\documentclass[a4paper, 11pt]{report}

\usepackage[margin=0.78in]{geometry}
\usepackage{float}
\usepackage{graphicx}
\graphicspath{{images/}}
\usepackage{epstopdf}
\usepackage{amsmath}
\usepackage{placeins}
\usepackage{bm} 
\usepackage{amsthm}

\usepackage{mathtools}
\usepackage{amssymb}
\usepackage[english]{babel}
\usepackage{graphicx}
\usepackage{sidecap}
\usepackage{caption}
\usepackage{multicol}
\usepackage{multirow}
\usepackage{tabularx}
\usepackage{wrapfig}
\usepackage[dvipsnames]{xcolor}
\usepackage[square, numbers, comma, sort&compress]{natbib}
\usepackage{hyperref}
\usepackage{nameref}
\definecolor{mygrayfonce}{gray}{0.4}

\title{Signal-to-noise ratio analysis of single-pixel detection multiplexing under photon-noise. \\ Cases of Hadamard and Cosine positive modulation.}

\usepackage{authblk}
\author[1]{Camille Scott\'e}
\author[1]{Fr{\'e}d{\'e}ric Galland}
\author[1]{Herv\'e Rigneault}

\affil[1]{Aix Marseille University, CNRS, Centrale Marseille, Institut Fresnel, Marseille, France}

\begin{document}

\maketitle

\section*{}

\textbf{Abstract}: In typical single-pixel detection multiplexing, an unknown object is sequentially illuminated with intensity patterns: the total signal is summed into a single-pixel detector and is then demultiplexed to retrieve the object. Because of measurement noise, the retrieved object differs from the ground truth by some error quantified by the signal-to-noise ratio (SNR). In situations where the noise only arises from the photon counting process, it has not been made clear if single-pixel detection multiplexing leads to a better SNR than simply scanning the object with a focused intensity spot - a modality known as raster scanning. This study theoretically assesses the SNR associated with certain types of single-pixel detection multiplexing, and compares it with raster scanning. In particular, we show that, under photon noise, when the positive intensity modulation is based on Hadamard or Cosine patterns, single-pixel detection multiplexing does not systematically improve the SNR as compared to raster scanning. Instead, it only improves the SNR on object pixels at least $k$ times brighter than the object mean signal $\bar{x}$, where $k$ is a constant that depends on the modulation scheme. 

\section*{Introduction}
\label{sec: SI_Intro}

\vspace{1cm}


\noindent The aim of the study is to compare the signal-to-noise ratio associated with raster-scanning and certain categories of single-pixel detection multiplexing, when the noise only arises from the photon-counting process. 

\begin{wrapfigure}{l}{0.68\textwidth}
  \begin{center}
    \includegraphics[width=0.65\textwidth]{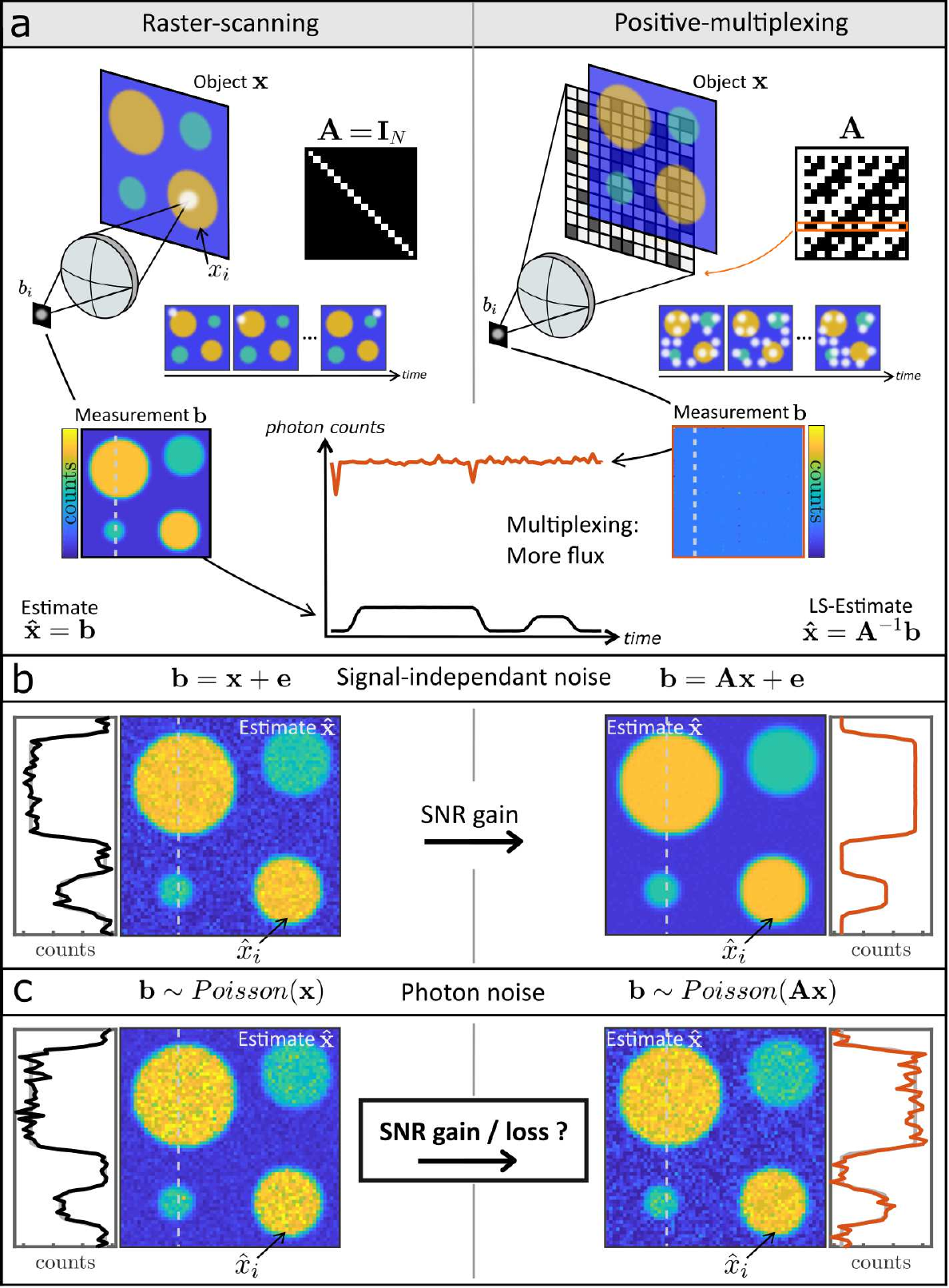} 
    \caption{(a) Schematic representation of the forward model for raster-scanning and positive-multiplexing in the absence of noise. (b) Example of estimated object in the presence of additive-white Gaussian noise (signal independent noise). (c) Example of estimated object in the presence of photon noise. $\mathbf{x}$: intensity object, $\mathbf{\hat{x}}$: estimation of $\mathbf{x}$, $\mathbf{b}$: measurement, $\mathbf{A}$: invertible multiplexing matrix, $\mathbf{I}_N$: identity matrix, LS : least-square.}
  \end{center}
  \label{fig:dessin_intro}
\end{wrapfigure}



\noindent \hspace{5cm} \\
\noindent In raster scanning (Fig.~1 a-left), an unknown object $\mathbf{x}$ is scanned point by point-by-point with a focused intensity spot. In a single-pixel camera (Fig.~1 a-right), the object is sequentially illuminated by intensity patterns: the total reflected or transmitted intensity is summed into a single-pixel detector and must be demultiplexed to retrieve the object \cite{Edgar2019}. This intensity modulation multiplexing is referred to as \textit{Positive-multiplexing}\footnote{because this work is also valid in the particular case of interferometric measurements where the object of interest is the field power spectrum, as will be shown in \ref{sec:Fourier}}. In both cases, because of measurement noise, the retrieved object $\mathbf{\hat{x}}$ differs from the ground truth object $\mathbf{x}$ by some error (Fig.~1 b-c). This error is quantified by the signal-to-noise ratio (SNR), or equivalently by the mean-square error (MSE). When the noise arises from the detector electronics (Fig.~1 b), it is well known common positive-multiplexing types dramatically improve the SNR as compared to raster-scanning \cite{Fellgett1967, DeVerse2000}. Yet, when the noise only arises from the photon counting process (Fig.~1 c), the answer has not been made clear, despite the development of high-performance single-pixel detectors that tends to make measurements more and more likely to be limited by photon noise only.
\\ \\ \\ \\ \\


\noindent The present document assesses the \textit{theoretical} SNR performances of raster-scanning and positive-multiplexing. We therefore consider a general intensity object $\mathbf{x}$ that we seek to estimate with least-square estimation (LS). We derive the theoretical SNR associated with raster-scanning and positive-multiplexing, for three measurements strategies, both for a general matrix and for Hadamard-based and Cosine-based positive-multiplexing. \\

\noindent The document is organised as followed: In Part 1, we define the assumptions and framework of the study. In Part 2, we derive the general expressions for the MSE (or SNR) associated to raster-scanning and positive-multiplexing. In Part 3, we derive the MSE for three common positive-multiplexing schemes. In Part 4, we apply these results to positive-multiplexing based on Hadamard matrices and to modulation with positive cosines. We find that for these classes of positive-multiplexing, the MSE is constant on most object pixels, and depends on the average signal contained in the object. In Part 5, we study the robustness of raster-scanning and positive-multiplexing to several perturbations. Last, in Part 6, we elaborate on some general properties of positive-multiplexing matrices that lead to a constant MSE. 

{\renewcommand{\arraystretch}{2}
\begin{table} [H]
\centering
\begin{tabularx}{\linewidth}{|p{5cm}|X|}
\hline
SNR dependence & Studied case \\
\hline
Object $\mathbf{x}$ & General object $\mathbf{x}$ \\
\hline
Estimation method & Least-square (LS) \\
\hline
Multiplexing scheme & One-step, Two-step, Dual-detection \\
\hline
Multiplexing matrix $\mathbf{A}$  & General matrix $\mathbf{A}$ ; positive-Hadamard ($\mathbf{H_1}$-matrix and $\mathbf{S}$-matrix), positive-Cosines \\
\hline
Noise \& measurement model & Photon noise, perturbation of the model with various noise sources or nuisances \\
\hline
\end{tabularx}
\caption{Overview of the parameters influencing the SNR assessed in this study}
\label{tab:SumUp_Framework}
\end{table}

\noindent The dependence of the SNR on other parameters, such as the object signal itself or the estimation strategy will assessed numerically and experimentally in a forthcoming publication, in order to provide implementable user guidelines.

\tableofcontents


\chapter*{Notations and properties}
\label{sec: SI_Notations and properties}


\begin{itemize}
\item $\mathbf{I}_N$ : $N \times N$ identity matrix
\item $\mathbf{J}_N$ : $N \times N$ matrix of ones
\item $\mathbf{1}_N$ : $N \times 1$ vector of ones
\item $\mathbf{e_i}$ : $N \times 1$ vector of zeros with a one at index $i$
\item $\mathbf{1e_1}^T$ : matrix with ones on the first column and zeros elsewhere
\item $\mathbf{e_1 1}^T$ : matrix with ones the first row and zeros elsewhere
\item $\odot$ : Point-wise product
\item $\otimes$ : Kronecker product
\item $\scriptstyle T$ : Transpose 
\item $*$ : Conjugate
\item $\scriptstyle H$ : Conjugate transpose
\item $\scriptstyle +$ : Pseudo-inverse
\item $\hat{\mathbf{x}}$ : estimate of $\mathbf{x}$
\item $\bar{x}$ : average over all pixels of $\mathbf{x}$
\item $diag(\mathbf{A})$ : builds the vector from the diagonal elements of $\mathbf{A}$
\item $Diag(\mathbf{x})$ : builds the diagonal matrix which elements are made of the vector $\mathbf{x}$
\item $vec(\mathbf{A})$ : stacks the columns of $\mathbf{A}$ into a vector.
\end{itemize}

In general $\mathbf{A}$ denotes a matrix ,  $\mathbf{a}$ a vector and $a$ a scalar (Except for the variance denoted $\mathbf{V}$).\\

We also make extensive use of the following properties \cite{MatrixCookBook2012}: 
\begin{equation}
diag(\mathbf{A} Diag(\mathbf{b})\mathbf{C}) = (\mathbf{A} \odot \mathbf{C}^T)  \mathbf{b} \hspace{0.2cm} \text{(for real quantities)}
\label{eqn:diag_property}
\end{equation}
\begin{equation}
diag(\mathbf{A} Diag(\mathbf{b})\mathbf{C}) = (\mathbf{A} \odot \mathbf{C}^*)  \mathbf{b} \hspace{0.2cm} \text{(for complex quantities)}
\label{eqn:diag_propertyCmplx}
\end{equation}
\begin{equation}
(\mathbf{A}+\mathbf{b}\mathbf{c}^T)^{-1} = \mathbf{A}^{-1} - \frac{\mathbf{A}^{-1}\mathbf{b}\mathbf{c}^T\mathbf{A}^{-1}}{1+\mathbf{c}^T \mathbf{A}^{-1} \mathbf{b}}
\label{eqn:Sherman_Morrison}
\end{equation}
\begin{equation}
vec(\mathbf{A} \mathbf{X} \mathbf{B}) =  (\mathbf{B}^T \otimes \mathbf{A}) vec(\mathbf{X})
\label{eqn:Vec_pty}
\end{equation}

\chapter{Framework of the study}
\label{sec: SI_Model}

To begin with, we introduce the model with its associated assumptions and figures-of-merit. 

\section{General model}
\label{sec: Model_definitions}
\label{sec: General_model}

We consider the following simple linear forward model: a real positive object $\mathbf{x}$ is measured through a matrix $\mathbf{A}$, leading to an ideal noiseless measurement vector $\mathbf{b_0}$:
\begin{equation}
\mathbf{b_0} = \mathbf{Ax}
\label{eqn:noiseless_meas}
\end{equation}
where $\mathbf{b_0} \in \mathbb{R}^N_+$ is the vector of $N$ noiseless measures, $\mathbf{x} \in \mathbb{R}^N_+$ is the object of interest, and $\mathbf{A}\in \mathbb{R}_{+}^{N \times N}$ is the $N \times N$ measurement matrix, made of positive coefficients $a_{ij} \geq 0$. In multiplexing, $\mathbf{A}$ is built with some mixing coefficients, while in raster-scanning, it is simply equal to the identity matrix $\mathbf{I}_N$.
When the noise only arises from the photon-counting process, the noisy measurements read: 
\begin{equation} 
\boxed{\mathbf{b} \sim Poisson(\mathbf{Ax})}
\label{eqn:completemodel}
\end{equation}
where $\mathbf{b} \in \mathbb{N}^N$ is the vector of observed photon counts. Each measured number of photons $b_i$ is a random variable whose probability law is assumed to be a Poisson distribution of mean $\langle b_i \rangle = \langle \delta b_i^2 \rangle = b_{0i}$, where $\delta b_i = b_{0i} - b_i$ is the measurement error.

\section{Figures-of-merit: SNR and MSE}
\label{sec: Figures-of-merit}
Since the measurements $\mathbf{b}$ are noisy, one cannot perfectly access the ground-truth object $\mathbf{x}$ but can only estimate it. This estimate, denoted $\hat{\mathbf{x}}$ , is directly equal to the measurements for raster-scanning, and to their demodulation for multiplexing. In both cases, it differs from $\mathbf{x}$ by some error $\delta\hat{\mathbf{x}} = \hat{\mathbf{x}} - \mathbf{x}$. The aim of this work is to assess which of raster-scanning or positive-multiplexing leads to the smallest error.
This is assessed via the mean-square error (MSE) and signal-to-noise ratio (SNR). Both inform on how precise and accurate is the estimate $\hat{\mathbf{x}}$  on each object pixel $i$:  
\begin{equation}
\boxed{MSE(\hat{x}_i) = \langle(\hat{x}_i - x_i)^2 \rangle \hspace{0.3cm} \text{;} \hspace{0.3cm} SNR(\hat{x}_i) = \frac{x_i}{\sqrt{MSE(\hat{x}_i)}} }
\label{eqn:def_mse} 
\end{equation}
The potential SNR gain or loss brought by multiplexing over raster-scanning can be quantified with the following ratio:
\begin{equation} \label{eqn:def_G}
G_i = \frac{SNR(\hat{x}_i)_{multiplex}}{SNR(\hat{x}_i)_{raster-scan}} = \sqrt{\frac{MSE(\hat{x}_i)_{raster-scan}}{MSE(\hat{x}_i)_{multiplex}}}
\end{equation}
If $G_{i}>1$, multiplexing improves the SNR on pixel $i$, as compared to raster-scanning, and conversely.\\
Since the SNR and MSE are directly related, and to bypass the additional dependence on the object ground-truth, in the following we only give results in terms of MSE. The rest of the present document focuses on deriving the MSE analytical expressions associated with positive-multiplexing in different scenarios, and compare them with the MSE associated with raster-scanning.  

\section{Framework and assumptions}
\label{sec: Further assumptions}
Comparing the SNR performances of multiplexing and raster-scanning could be done for a multitude of scenarios. We choose to restrict our study to the following framework: 
\begin{itemize}
\item The quantities to estimate $\mathbf{x}$ are real and positive 
\item The detected quantities $\mathbf{b}$ are  positive intensities (or photon counts)
\item The multiplexing matrix $\mathbf{A}$ is real and positive
\item Although we mostly focus on cases in which multiplexing is achieved via intensity modulation (i.e. when the measurement is an incoherent sum of intensities), the study also holds for the particular case of interferometric measurements where the object of interest is the field power spectrum (e.g. Fourier-transform infrared spectroscopy), see section \ref{sec:Fourier}. 
These four first assumptions define what we call \textit{Positive-multiplexing}
\item The considered systems are linear systems
\item The measurements are assumed to be statistically independent
\item The system resolution is infinitely smaller than the finest details of the considered structures 
\item We work at fixed sampling: the number of measurements equals the number of probed object pixels $N$. The multiplexing matrix $\mathbf{A}$ is thus square. It is also assumed to be invertible 
\item In order to derive analytical expressions, in this document we perform all estimations with the least-square estimator. The effect estimators with positivity constrained adapted to photon-noise are studied in the article main text. 
\item Unless otherwise stated, the results are valid for any object dimensionality (1-D, 2-D, ...), as long as the variables can be rearranged in the form of equation \eqref{eqn:noiseless_meas}. 
\end{itemize}

\noindent Last, except in the dedicated section \ref{sec:SI_constantNbPhotons}, the comparison between raster-scanning and positive-multiplexing is \textit{not} performed at constant photon number. Rather, we compare raster scanning and positive-multiplexing for fixed exposure time and irradiance. On the example of Fig.1, this means each sample pixel is illuminated with the same light power: if in raster-scanning, each pixel of the object is illuminated with 1 mW during 1 ms, then in positive-multiplexing, each pixel of the object will also see 1 mW of incident light during 1 ms. This results in a consequently higher measured number of photons for positive-multiplexing (Fig.~1 a).
As we will see, even in this advantageous case, positive-multiplexing does not always lead to a better SNR than raster-scanning. 
%

\chapter{Results for a general matrix $\mathbf{A}$}
\label{sec: SI_Model}

In this section we derive the theoretical expressions for the MSE associated with raster-scanning and positive-multiplexing, from which the SNR expressions can be deduced using equation~\eqref{eqn:def_mse}. \\
For positive-multiplexing, MSE depends on the estimation method employed to demodulate the measurements. In this document, we use the least-square estimation (LS). 

\section{Least-square estimation (LS)}
\label{sec: LS estimator}
The LS estimator minimizes the squared $l_2$ norm between the noisy and noiseless measurements: 
\begin{equation} 
\mathbf{\hat{x}} = argmin ||\mathbf{b}-\mathbf{b_0}||^2
\end{equation}
If $\mathbf{A}$ is invertible, the LS solution reads:
\begin{equation} 
\mathbf{\hat{x}} = \mathbf{A}^{-1} \mathbf{b}
\label{eqn:LSsol}
\end{equation}
This estimation is unbiased: 
\begin{equation}
\langle \mathbf{\hat{x}} \rangle = \mathbf{A^{-1}} \langle \mathbf{b} \rangle = \mathbf{A^{-1}} \mathbf{b_0} = \mathbf{x}
\label{eqn:LS_unbiased}
\end{equation}
This estimator is optimal in the sense of the maximum-likelihood for additive-white Gaussian noise, but may not be optimal under photon-noise. 
In this document, we use the LS estimator to access an unbiased estimation and to derive the MSE.\\ analytical expressions. 

\noindent Since the LS estimator is unbiased, the following relation holds: 
\begin{equation}
\mathbf{MSE}(\hat{\mathbf{x}}) = \mathbf{V}(\hat{\mathbf{x}}) = diag (\mathbf{\Gamma}) 
\label{eqn:GeneralFormula_VarianceDiagCov}
\end{equation}
The MSE is equal to the estimation variance $\mathbf{V}(\hat{\mathbf{x}})$, which is itself equal to the diagonal of the covariance matrix $\mathbf{\Gamma} = \langle \mathbf{\delta\hat{x}}\mathbf{\delta\hat{x}}^T  \rangle$. 

\section{MSE for positive-multiplexing}
\label{sec: MSE_expression_LS}

We remind that the measurements read: 
\begin{equation} 
\mathbf{b} \sim Poisson(\mathbf{Ax}) 
\label{eqn:completemodel}
\end{equation}
The estimation error arising when estimating $\mathbf{\hat{x}}$ with LS via equation \eqref{eqn:LSsol} is: 
\begin{equation}
\mathbf{\delta\hat{x}} = \mathbf{\hat{x}} - \mathbf{x} = \mathbf{A^{-1}}(\mathbf{b}-\mathbf{b_0}) = \mathbf{A^{-1}}\delta\mathbf{b}
\label{eqn:ErrorLS}
\end{equation}
Inserting this in equation \eqref{eqn:GeneralFormula_VarianceDiagCov} yields:
\begin{equation} 
\mathbf{\Gamma} = \langle \delta\hat{\mathbf{x}}\delta\hat{\mathbf{x}}^{T}\rangle = \mathbf{A}^{-1} \langle \delta\mathbf{b}\delta\mathbf{b}^{T} \rangle  \mathbf{A}^{-T} 
\end{equation}
Since the measurements are assumed to be statistically independent and to follow a Poisson distribution,  $\langle \delta\mathbf{b}\delta\mathbf{b}^{T}\rangle $ is a diagonal matrix with elements $\langle \delta b_i^2 \rangle = \langle b_i \rangle = b_{0i} $ (Poisson distribution properties). Thus: 
\begin{equation}
\mathbf{\Gamma} 
=  \mathbf{A}^{-1} Diag(\mathbf{b_0}) \mathbf{A}^{-T}
=  \mathbf{A}^{-1} Diag(\mathbf{A} \mathbf{x}) \mathbf{A}^{-T} 
\label{eqn:cov_poiss}
\end{equation}
The MSE and estimation variance are finally obtained by selecting the diagonal of $\mathbf{\Gamma}$ (using equations \eqref{eqn:GeneralFormula_VarianceDiagCov} and \eqref{eqn:diag_property}). Hence, the MSE associated with positive-multiplexing reads:  
\begin{equation}
\boxed{\mathbf{MSE}(\hat{\mathbf{x}}) = \mathbf{V}(\mathbf{\hat{x}}) = (\mathbf{A}^{-1} \odot \mathbf{A}^{-1})\mathbf{A}\mathbf{x}}
\label{eqn:VarianceGeneralFormula_Real}
\end{equation}

\section{MSE for raster-scanning}

\label{sec: Raster-scanning}

In raster-scanning (Fig.~1 a), the object $\mathbf{x}$ is probed point-by-point: each measurement $b_i$ relates to one object pixel intensity $x_i$. Although in practice there is no multiplexing matrix, the latter can be viewed as the identity matrix $\mathbf{A} = \mathbf{I}_N$. The measurements read: 
\begin{equation} 
\mathbf{b} \sim Poisson(\mathbf{x})
\end{equation}
The object estimate is directly equal to the noisy measurement: 
\begin{equation}
\mathbf{\hat{x}} = \mathbf{b}
\end{equation}
and the error is $\mathbf{\delta\hat{x}} = \delta\mathbf{b}$. 
Applying \eqref{eqn:VarianceGeneralFormula_Real} with $\mathbf{A} = \mathbf{I}_N$ gives:
\begin{equation}
\boxed{\mathbf{MSE}(\hat{\mathbf{x}}) = \mathbf{V}(\hat{\mathbf{x}})= \mathbf{x}}
\label{eqn:Variance_RS}
\end{equation}
In other words, in raster-scanning, the MSE (or variance) is equal to the object pixel intensity on every pixel $i$. Therefore, from equation \eqref{eqn:def_mse}, in raster-scanning the SNR reads $\mathbf{SNR}(\hat{\mathbf{x}}) = \sqrt{\mathbf{x}}$. \\ \\

\noindent In the following, we only use LS-estimation, therefore we mostly speak about variance since:
\begin{equation}
\mathbf{MSE}(\hat{\mathbf{x}}) = \mathbf{V(\hat{\mathbf{x}})}
\label{eqn:MSEequalVariance}
\end{equation}
which is related to the SNR via equation \eqref{eqn:def_mse}.

\chapter{Results for three positive-multiplexing schemes}
\label{sec: SI_Modalities}
\label{sec: SI_Modalities_Schemes}
In practise, positive-multiplexing can be implemented in many different ways, which may impact the final SNR. 
In this section, we detail three possible positive-multiplexing schemes. We associate each scheme with its equivalent multiplexing matrix and theoretical variance. The figure here-in-below encompasses the main results of this section. Explanations and details are given in the text which follows. 
\label{sec: Modalities_and_variance}
\begin{figure} [H]
\centering
\includegraphics[width=\linewidth]{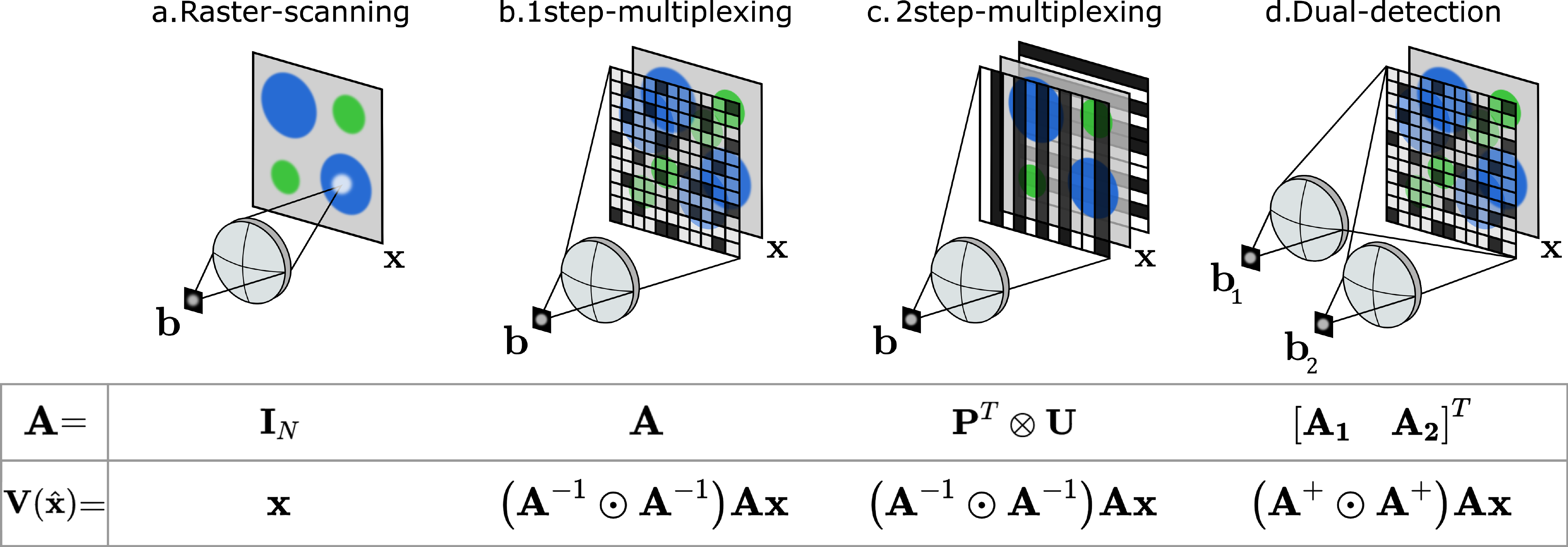} 
\caption {Schematic of 2-D multiplexing for raster-scanning and for the considered multiplexing modalities. To each scheme corresponds an equivalent matrix $\mathbf{A}$ and estimation variance $\mathbf{V}(\hat{\mathbf{x}})$ ($=\mathbf{MSE}(\hat{\mathbf{x}})$). All details are given in the text. (b) In one-step multiplexing, the patterns derive from the rows of $\mathbf{A}$. (c) Two-step multiplexing involves two independent sets of patterns: one set derives from the rows of a matrix $\mathbf{U}$ and the other from the rows of a matrix $\mathbf{P}$ (see Fig.~\ref{fig:dessin_Modalities} c). (d) Dual-detection involves two complementary measurements. The equivalent matrix is a bloc matrix (equation \eqref{eqn:modelBalancedBlock}). In this scheme, it is sometimes possible to substract the two measurements but this leads to a different model and theoretical variance (section~\ref{sec:noteBalancedDet}). 
}
\label{fig:dessin_Modalities_General}
\end{figure}

\section{One-step multiplexing}
\label{sec: One-step multiplexing}
In one-step multiplexing (Fig.~\ref{fig:dessin_Modalities_General} b), the object is probed with patterns corresponding to its dimensionality. As detailed in Fig.~\ref{fig:dessin_Modalities} (a-b), each measurement $b_i$ is the sum of the point-wise product between the object and a pattern that derives from the $i^{th}$ row of the multiplexing matrix $\mathbf{A}$. The variance (or MSE) is the same as in equation \eqref{eqn:VarianceGeneralFormula_Real}: 
\begin{equation}
\boxed{\mathbf{V}(\mathbf{\hat{x}}) = (\mathbf{A}^{-1} \odot \mathbf{A}^{-1})\mathbf{A}\mathbf{x}}
\label{eqn:OneStep_GeneralFormula}
\end{equation}


\section{Two-step multiplexing}
\label{sec: Two-step multiplexing}
If the object has more than one dimension, it is possible to perform multiplexing in one single step (as above), or in several steps, if its dimensions are separable. In the case of a 2-D object, this would involve two independent 1-D multiplexing stages that probe uncorrelated dimensions of the object, such as the vertical and horizontal dimensions of a 2-D spatial object (Fig.~\ref{fig:dessin_Modalities_General} c), or the spatial and spectral dimensions of a spatio-spectral object. 
Then, instead of multiplexing with a matrix $\mathbf{A}$ of dimension $N \times N$, multiplexing is performed independently by two matrices $\mathbf{P}$ and $\mathbf{U}$, each of size $\sqrt{N} \times \sqrt{N}$, as detailed in Fig.~\ref{fig:dessin_Modalities}(c). Writing $\mathbf{x} = vec(\mathbf{X})$, and using \eqref{eqn:Vec_pty} allows writing the measure as:
\begin{equation}
\mathbf{b_0} = vec(\mathbf{U} \mathbf{X} \mathbf{P}) =  (\mathbf{P}^T \otimes \mathbf{U}) \mathbf{x}  =\mathbf{A} \mathbf{x} 
\label{eqn:noiseless_meas_Kron}
\end{equation} 
and 
\begin{equation}
\mathbf{b} \sim Poisson \left( ( \mathbf{P}^T \otimes \mathbf{U}) \mathbf{x} \right) = Poisson ( \mathbf{A} \mathbf{x} )
\end{equation} 
Hence, this is equivalent to multiplexing with an equivalent matrix: 
\begin{equation}
\mathbf{A} = \mathbf{P}^T \otimes \mathbf{U}
\end{equation}
leading to an equivalent variance derived from equation \eqref{eqn:VarianceGeneralFormula_Real}: 
\begin{equation} 
\boxed{ \begin{split}
\mathbf{V} (\mathbf{\hat{x}}) & = \left( (\mathbf{P}^T \otimes \mathbf{U})^{-1} \odot (\mathbf{P}^T \otimes \mathbf{U})^{-1} \right)(\mathbf{P}^T \otimes \mathbf{U})\mathbf{x} \\
& = \left( (\mathbf{P}^{-T} \odot \mathbf{P}^{-T}) \mathbf{P}^T  \otimes ( \mathbf{U}^{-1} \odot \mathbf{U}^{-1} ) \mathbf{U} \right) \mathbf{x}  \\
\end{split} }
\label{eqn:Kron_GeneralFormula}
\end{equation}

\begin{figure} [H]
\centering
\includegraphics[width=\linewidth]{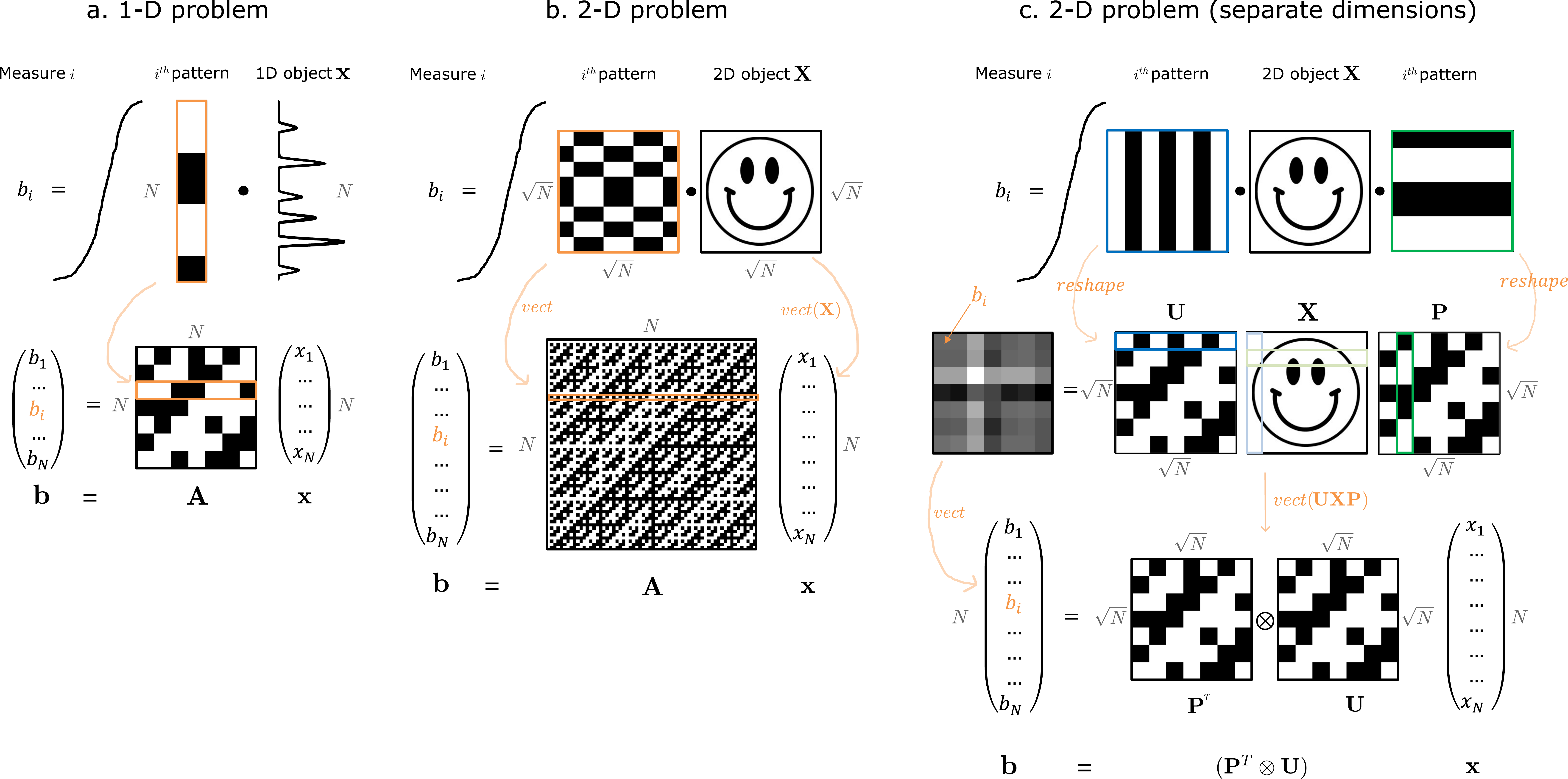} 
\caption {(a) One-step multiplexing in 1-D: the $i^{th}$ row of $\mathbf{A}$ is directly the $i^{th}$ pattern (b) One-step multiplexing in 2-D: the $i^{th}$ row of $\mathbf{A}$ is reshaped to obtain a a 2-D pattern. (c) Two-step multiplexing: if the object dimensions are separable and probed independently, multiplexing in 2-D can be performed in 2-steps, what we call two-step multiplexing.}
\label{fig:dessin_Modalities}
\end{figure}

\section{Dual-detection}
\label{sec: Balanced detection}
A dual detection scheme (Fig.~\ref{fig:dessin_Modalities_General} d) involves two complementary measurements $\mathbf{b_1}$ and $\mathbf{b_2}$. For example, when multiplexing 0 and 1 entries, the signal corresponding to zeros not collected by the first detector is collected by the second detector. This is equivalent to associating a multiplexing matrix $\mathbf{A_1}$ to  the measurements $\mathbf{b_1}$ and multiplexing matrix $\mathbf{A_2}$ to the measurements $\mathbf{b_2}$, such that:
\begin{equation}
\mathbf{A_1} + \mathbf{A_2} = \mathbf{J_N} \\ 
\label{itemlist:balanceddet_def}
\end{equation}

\noindent The measurements $\mathbf{b_1}$ and $\mathbf{b_2}$ associated with matrices $\mathbf{A_1}$ and $\mathbf{A_2}$ respectively read:
\begin{equation}
\left\{
    \begin{array}{ll}
        \mathbf{b_1} \sim Poisson(\mathbf{A_1} \mathbf{x})  \\
        \mathbf{b_2} \sim Poisson(\mathbf{A_2} \mathbf{x})
    \end{array}
\right.
\end{equation}
The measurement model is obtained by combining the two measurements into the same vector: 
\begin{equation}
\mathbf{b} =
\begin{bmatrix}
\mathbf{b_1}\\
\mathbf{b_2}
\end{bmatrix} 
\sim Poisson \left( \begin{bmatrix}
\mathbf{A_1}\\
\mathbf{A_2}
\end{bmatrix} 
\mathbf{x} \right)
= 
Poisson (\mathbf{A} \mathbf{x})
\label{eqn:modelBalancedBlock}
\end{equation}
where $\mathbf{A}$ is a block matrix of size $2N \times N$. It can be seen as an equivalent multiplexing matrix:
\begin{equation}
\mathbf{A} =  \begin{bmatrix}
\mathbf{A_1} \hspace{0.1cm}  \mathbf{A_2}
\end{bmatrix} ^T
\end{equation}
Then, if $\mathbf{A}^{T} \mathbf{A}$ is not singular, $mathbf{x}$ can be estimated via $\hat{mathbf{x}} = \mathbf{A}^{+} \mathbf{b}$, where $\mathbf{A}^{+} $ is the pseudo-inverse: $
\mathbf{A}^{+} =  (\mathbf{A}^{T} \mathbf{A})^{-1} \mathbf{A}^{T}$.
Replacing in equation \eqref{eqn:VarianceGeneralFormula_Real} leads to the associated estimation variance:
\begin{equation}
\boxed{\mathbf{V}(\hat{\mathbf{x}}) = (\mathbf{A}^{+} \odot \mathbf{A}^{+}) \mathbf{A} \mathbf{x}}
\label{eqn:VarianceGeneralBalanced_Block}
\end{equation}

\subsubsection*{Alternative approach: the balanced approach}
\label{sec:noteBalancedDet}
In the literature (eg. \cite{Rodriguez2016, Soldevila2016, Zhang2015, Zhang2020}), it is often found that the reconstituted measure $\mathbf{b}$ is rather expressed as: 
\begin{equation} 
\mathbf{b} = \mathbf{b_1} - \mathbf{b_2} \sim Poisson(\mathbf{A_1 x}) - Poisson(\mathbf{A_2 x}) \neq Poisson(\mathbf{A x})
\label{eqn:modelBalancedSubtract}
\end{equation}
Therefore, it does not follow the prerequisite measurement model of equation \eqref{eqn:completemodel}. 
Still, it is possible to define an equivalent
multiplexing matrix $\mathbf{A}$: 
\begin{equation}
\mathbf{A} = \mathbf{A_1} - \mathbf{A_2}
\end{equation} 
and to perform the estimation through: $\mathbf{\hat{x}} = \mathbf{A}^{-1} \mathbf{b}$, if $\mathbf{A}$ is invertible. 
This is the strategy used in the literature cited above. 
Such an estimation strategy may seem surprising since it may lead to a significant information loss. 
Yet, \textit{in some particular cases}, this strategy is justified because it is computationally efficient and leads to approximately the same variance than when using the model of equation \eqref{eqn:modelBalancedBlock} (see section~\ref{sec:JustificationBalancedDetection} for more details). To derive the variance associated with this balanced approach, one cannot directly apply equation \eqref{eqn:VarianceGeneralFormula_Real}. Rather, one needs to go back to the definition of the covariance matrix, and finds: 
\begin{equation}
\boxed{\mathbf{V}(\hat{\mathbf{x}}) =  N\bar{x}(\mathbf{A}^{-1} \odot \mathbf{A}^{-1}) \mathbf{1}_N}
\label{eqn:Balanced_GeneralVarReal}
\end{equation}
\underline{Proof}: 
Covariance matrix:
\begin{equation}
\begin{split}
\mathbf{\Gamma} & \equiv  \langle \delta\mathbf{\hat{x}}\delta\mathbf{\hat{x}}^{T} \rangle\\
& =  \mathbf{A}^{-1} \langle \delta\mathbf{b}\delta\mathbf{b}^T\rangle  \mathbf{A}^{-T} \\
& =  \mathbf{A}^{-1} \langle (\mathbf{b}-\mathbf{b_0})(\mathbf{b}-\mathbf{b_0})^T\rangle  \mathbf{A}^{-T} \\
& =  \mathbf{A}^{-1} \langle ( \delta \mathbf{b_1} -\delta \mathbf{b_2} )( \delta \mathbf{b_1} -\delta \mathbf{b_2} )^T \rangle  \mathbf{A}^{-T} \\
\end{split}
\end{equation}
Since the measurements are statistically independent, $\langle \delta \mathbf{b_1} \delta \mathbf{b_2}^T \rangle = \langle \delta \mathbf{b_2} \delta \mathbf{b_1}^T \rangle = 0$, and: 
\begin{equation}
\begin{split}
\mathbf{\Gamma} & =  \mathbf{A}^{-1} ( \langle \delta \mathbf{b_1} \delta \mathbf{b_1}^T \rangle + \langle \delta \mathbf{b_2} \delta \mathbf{b_2}^T \rangle )\mathbf{A}^{-T} \\
& =  \mathbf{A}^{-1} Diag(\mathbf{b_{1,0}}+\mathbf{b_{2,0}}) \mathbf{A}^{-T} \\
&=  \mathbf{A}^{-1} Diag((\mathbf{A_1}+\mathbf{A_2}) \mathbf{x}) \mathbf{A}^{-T} \\
&=  \mathbf{A}^{-1} Diag(\mathbf{J_N}\mathbf{x}) \mathbf{A}^{-T} =  N \bar{x} (\mathbf{A}^{T} \mathbf{A} )^{-1} 
\end{split}
\label{eqn:Cov_Balanced_General}
\end{equation}
Using equations \eqref{eqn:GeneralFormula_VarianceDiagCov} and \eqref{eqn:diag_property} leads to the variance expression of equation~\eqref{eqn:Balanced_GeneralVarReal}.

\chapter{Properties of positive-Hadamard-based multiplexing \& positive-Cosine multiplexing}
\label{sec: SI_CasesHadaFourier}

So far, we have derived the expressions of the MSE for raster-scanning and three positive-multiplexing schemes, for a general multiplexing matrix $\mathbf{A}$ verifying several assumptions (section~\ref{sec: Further assumptions}). In the present section, we derive the MSE for two widely implemented positive-multiplexing types, based on (i) positive-Hadamard multiplexing and (ii) positive-Cosine multiplexing.  \\
On the one hand, we consider positive-multiplexing based on Hadamard matrices, i.e. on binary matrices derived from the Hadamard matrix $\mathbf{H}$. We consider two specific cases: the $\mathbf{S}$-matrix (section \ref{sec:Multiplexing with the S-matrix}) and the positive-Hadamard matrix $\mathbf{H_1}$ (section \ref{sec:HadaPositive}).
On the other hand , we consider positive-Cosine multiplexing, i.e. modulation with cosine waveforms (section \ref{sec:Fourier}). \\

{\renewcommand{\arraystretch}{2}
\begin{table} [H]
\centering
\begin{tabularx}{\linewidth}{|X|X|X|X|}
\hline
Multiplexing class & \multicolumn{2}{|c|}{positive-Hadamard-based} &  positive-Cosine \\
\hline
Associated matrix & $\mathbf{S}$-matrix & $\mathbf{H_1}$-matrix & $\mathbf{C_1}$-matrix \\
\hline
Short description & modified Hadamard matrix with binary coefficients ($0$ or $1$)& Hadamard matrix with binary coefficients ($0$ or $1$) & matrix with cosine waveforms \\
\hline
\end{tabularx}
\caption{Summary of the multiplexing classes considered in this section.}
\label{tab:SumUp_multiplexMatrices}
\end{table}


\noindent This section is organised as followed: (i)  main result;
(ii) positive-Hadamard-based multiplexing (results, matrices and proofs); (iii) positive-Cosine multiplexing (results, details, matrices and proofs); (iv) note on dual-detection; (v) few numerical simulations to illustrate the results.

\section{Main result}
\label{sec: VarianceResults_HadaFourier} 


\noindent We prove that for both positive-Hadamard-based multiplexing and positive-Cosine multiplexing and for three positive-multiplexing scheme, 
the estimation variance obtained with least-square estimation is constant over the estimated object $\hat{\mathbf{x}}$ on most pixels $i$. We show that it is proportional to the average signal contained in the object $\bar{x}$, on \textit{most} object pixels $i$:
\begin{equation}
\boxed{MSE(\hat{x}_i) = V(\hat{x}_i) \approx k \bar{x}} \text{ if } N \gg 1
\label{eqn:Rule of Thumb}
\end{equation}
where $N$ is the number of pixels and $k$ is a positive constant.
Since the variance associated with raster-scanning equals the object itself (equation \eqref{eqn:Variance_RS}), positive-multiplexing improves the SNR over raster-scanning by a factor (equation \eqref{eqn:def_G}) equal to:
\begin{equation}
G_i = \sqrt{\frac{x_i}{k\bar{x}}} 
\label{eqn:Gratio_fourierHada}
\end{equation}
This means that, under the assumptions considered in this work, positive-multiplexing brings an improvement over raster-scanning only on object pixels $i$ that verify: 
\begin{equation}
\boxed{x_i \geq k \bar{x}}
\label{eqn:Rule of Thumb}
\end{equation}
The direct consequence is that the considered multiplexing strategies \textit{do not} systematically bring an improvement over raster-scanning: it improves the SNR only on object pixels $i$ with an intensity greater than the threshold value $k \bar{x}$. This result is of primary importance when choosing an optical design or a measurement strategy. 
The constant $k$ depends on the multiplexing matrix, on multiplexing modality (one-step multiplexing, two-step multiplexing, dual detection), and on some specificities detailed below. \\
\noindent \underline{Note:} The matrices presented in this section are defined so that they are directly implementable on physical systems. This implies that they do not necessarily have the same matrix norm \footnote{Largest singular value}. To adjust the constant $k$ for matrices norms or for multiplicative constants, refer to section~\ref{sec:Perturb_Gamma}. \\

\section{Positive-Hadamard-based multiplexing}
\label{sec: Hadamard-based multiplexing}

First, we focus on positive-multiplexing based on Hadamard matrices, based on binary matrices derived from the Hadamard matrix $\mathbf{H}$. In practise, this Hadamard-based positive multiplexing is often implemented by modulating the light intensity with encoding patterns (e.g. with a light modulator device) \cite{Harwit1979, DeVerse2000, Studer2012, Berto2017, Scotte2020, ScotteThesis2020}, such as in Fig.~1 a.
We consider two multiplexing matrices that are both based on Hadamard-multiplexing: the $\mathbf{S}$-matrix (section \ref{sec:Multiplexing with the S-matrix}) and the positive-Hadamard matrix $\mathbf{H_1}$ (section \ref{sec:HadaPositive}). 

\subsection{Definition and properties of the matrices $\mathbf{S}$ and $\mathbf{H_1}$}

Positive-multiplexing based on the Hadamard matrices makes use of modified Hadamard matrices such that the matrix entries are $0$ and $+1$ rather than $-1$ or $+1$. In this work, we consider :
\begin{itemize}
\item The $\mathbf{H_1}-$matrix (denoted as "positive-Hadamard matrix") is a binary Hadamard matrix where the $-1$ elements of $\mathbf{H}$ are replaced with $0$s. 
\item The $\mathbf{S}-$matrix is a binary matrix defined by \eqref{eqn:Smatrixpty}. 
A $\mathbf{S}-$matrix of size $N$ $\times$ $N$ can for instance be obtained by removing the first row and column of a $(N+1) \times (N+1)$ Hadamard matrix, and changing its $+1$s to $0$s and $-1$s to $+1$s.
\end{itemize}

\begin{figure} [h]
\centering
\textbf{\includegraphics[scale=0.7]{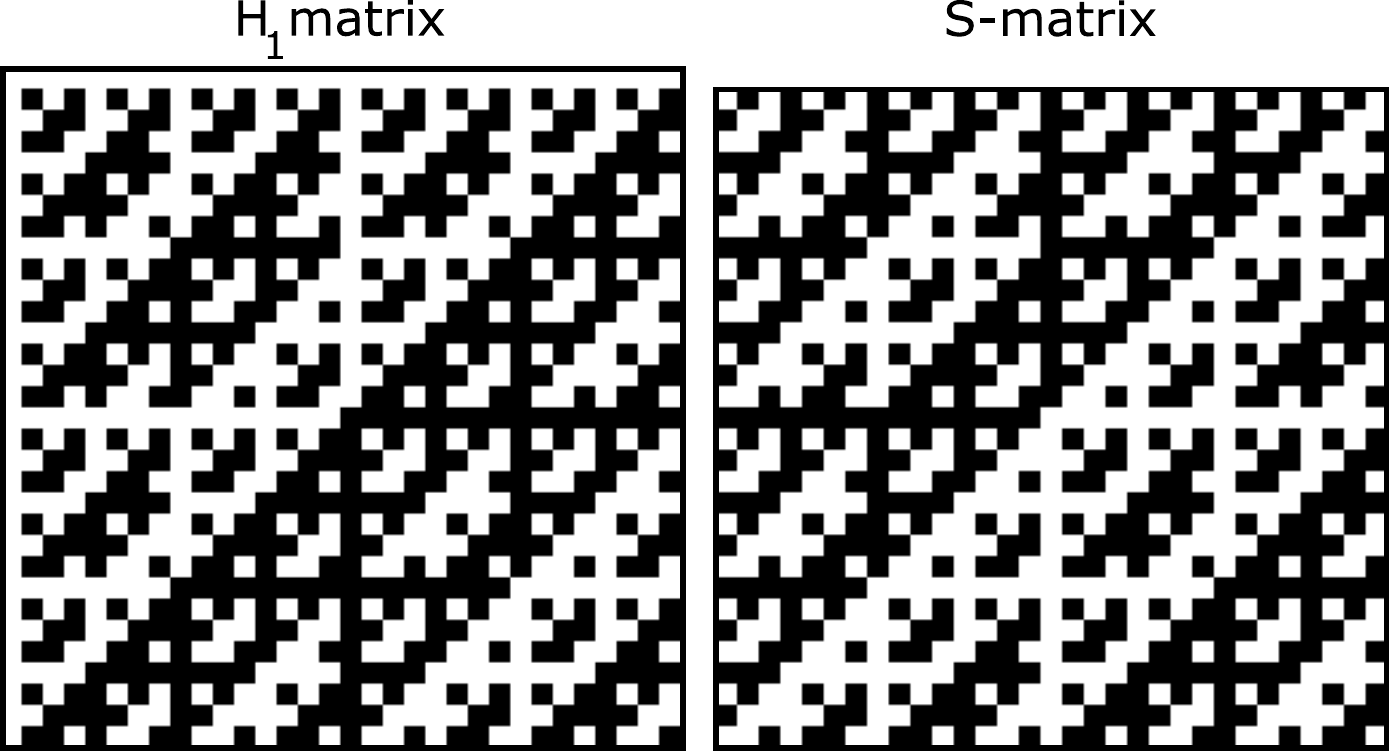} }
\caption{$\mathbf{H_1}$: Positive-Hadamard matrix (32$\times$32), and corresponding $\mathbf{S}-$-matrix (31$\times$31), with (0,1) $\equiv$ (black, white)}
\label{fig:SmatHada}
\end{figure}

\subsubsection*{The Hadamard matrix}
A Hadamard matrix $\mathbf{H}$ is a real square matrix with entries $-1$ and $1$ whose rows are pairwise orthogonal \cite{Sylvester1867, Hedayat1978, Harwit1979}. The orthogonality condition means that the dot product of any two distinct rows is zero; and it implies that, when comparing two rows, the number of matchings ($+1$) is equal to the number of mismatchings ($-1$).
A Hadamard matrix $\mathbf{H}$ of size $N \times N$ is such that \cite{Hedayat1978, Harwit1979}: 
\begin{equation}
\mathbf{HH}^T = \mathbf{H}^T \mathbf{H} = N \mathbf{I}_N
\label{eqn:Hadaproperty}
\end{equation}
Hadamard matrices do not exist for any $N$, but it is conjectured that there is at least one Hadamard matrix of order $N = 4p$ for every positive integer $p$ \cite{Wallis1976}. There are several subcategories of Hadamard matrices. In this work, we consider the widely used Sylvester type \cite{Sylvester1867}. For Sylvester Hadamard matrices, $N = 2^p$ and $\mathbf{H}$ is a symmetric matrix ($\mathbf{H}^T = \mathbf{H}$) in which each row and column contains the same number of $+1$ and $-1$ elements, except for the first row and column where all elements are $+1$s. \\

\noindent \underline{Note:} under additive white Gaussian noise (AWGN), Hadamard matrices provide an "optimal encoding design": among matrices which elements can be $-1$, $0$ or $1$, Hadamard matrices  minimize the MSE \cite{Hedayat1978, Harwit1979}.

\subsubsection*{The $\mathbf{S}-$matrix}
\label{sec:Some properties of the S-matrix}

A S-matrix of size $N$ $\times$ $N$ is a binary matrix defined by \cite{Harwit1979}: 
\begin{equation} 
\left\{
\begin{array}{ll}
\mathbf{S}^T\mathbf{S} = \mathbf{S}\mathbf{S}^T = \frac{N+1}{4}(\mathbf{I} + \mathbf{J}_N) \\
\mathbf{J}_N \mathbf{S} = \mathbf{S}\mathbf{J}_N = \frac{N+1}{2}\mathbf{J}_N
\end{array}
\right.
\label{eqn:Smatrixpty}
\end{equation}
This means that the sum of each column (or row) of a S-matrix is equal to $\frac{N+1}{2}$.
In this work, the S-matrix of size $N \times N$ is obtained by removing the first row and column of a $(N+1) \times (N+1)$ Sylvester Hadamard matrix, and changing its $+1$s to $0$s and $-1$s to $+1$s. In these conditions, $\mathbf{S}$ is of odd dimension, symmetric ($\mathbf{S} = \mathbf{S}^T$), and invertible with:
\begin{equation}
\mathbf{S}^{-1} = \frac{2}{N+1}(2 \mathbf{S}^T - \mathbf{J}_N)
\label{eqn:Sinv}
\end{equation}
and $\mathbf{S}^{-1}$ verifies:
\begin{equation}
\mathbf{S}^{-1} \odot \mathbf{S}^{-1} = \frac{4}{(N+1)^2}\mathbf{J}_N
\label{eqn:S.S}
\end{equation}
\underline{Note:} under additive white Gaussian noise, S-matrices minimize the MSE among matrices with entries $0$ and $1$ \cite{Harwit1979, Statistics1987, Drnovsek2013}.

\subsubsection*{The $\mathbf{H_1}-$matrix}

\label{sec:H1_Properties}
The $\mathbf{H_1}-$matrix can be expressed as a function of $\mathbf{H}$ and $\mathbf{J}_N$ (constant matrix made of $+1$ elements):
\begin{equation}
\mathbf{H_1} = \frac{1}{2}(\mathbf{J}_N + \mathbf{H} )
\label{eqn:def_H1}
\end{equation}
$\mathbf{H_1}$ is invertible, and its invert can be expressed analytically via equation \eqref{eqn:Sherman_Morrison}: 
\begin{equation}
\mathbf{H_1}^{-1} = \frac{2}{N}(\mathbf{H}^T - \frac{N}{2} \mathbf{e_1}\mathbf{e_1}^T)
\label{eqn:H1_invert}
\end{equation}
We denote $\mathbf{H_2}$ the complementary matrix of $\mathbf{H_1}$ ('negative' Hadamard matrix where the $+1$ elements of $\mathbf{H}$ are replaced with $0$s and $-1$ element with $+1$s): 
\begin{equation}
\mathbf{H_2} = \frac{1}{2}(\mathbf{J}_N - \mathbf{H})
\label{eqn:def_H2}
\end{equation}
$\mathbf{H_2}$ is not invertible. 
Note that, in addition to \eqref{eqn:Hadaproperty}, a Sylvester Hadamard matrix has, among others, the following properties: 
\begin{equation}
\mathbf{H} \odot \mathbf{H} = \mathbf{J}_N 
\label{eqn:H.H}
\end{equation}
\begin{equation}
\mathbf{J}_N \mathbf{H} = N \mathbf{1}\mathbf{e_1}^T
\label{eqn:JnH}
\end{equation}
\begin{equation}
\mathbf{H} \mathbf{J}_N  = N \mathbf{e_1} \mathbf{1}^T
\label{eqn:HJn}
\end{equation}
\begin{equation}
\mathbf{1}\mathbf{e_1}^T \mathbf{H} = \mathbf{J}_N 
\label{eqn:1e1H}
\end{equation}
\begin{equation}
\mathbf{H}\mathbf{1}\mathbf{e_1}^T  = N \mathbf{e_1} \mathbf{e_1}^T 
\label{eqn:H1e1}
\end{equation}

\subsection{MSE results}

For both matrices, the results of the associated estimation variances (or MSE), for LS-estimation, are given in the Tables below. The proofs are given in sections \ref{sec:Multiplexing with the S-matrix} and \ref{sec:HadaPositive}.

{\renewcommand{\arraystretch}{2}
\begin{table} [H]
\centering
\begin{tabularx}{\linewidth}{|p{1.8cm}|p{4.4cm}|p{4.4cm}|X|X|}
\hline
 &   One-step multiplexing & Two-step multiplexing  & \multicolumn{2}{|c|}{Dual-detection}  \\
\hline
$\mathbf{A}-$matrix &  $\mathbf{S}$ & $\mathbf{S} \otimes \mathbf{S}$  & $ \left[ \mathbf{S} \hspace{0.1cm} \mathbf{S_2} \right]^T$ & $\mathbf{S}-\mathbf{S_2}$ \\ 
\hline

\scalebox{1.2}{$V(\hat{x}_i)$}  & \scalebox{1.2}{$2 \bar{x}$} \hspace{0.1cm} $\scriptstyle (\forall i )$ & \scalebox{1.2}{$4 \bar{x}$} \hspace{0.1cm} $\scriptstyle (\forall i )$  & \scalebox{1.2}{$ \bar{x}$} \hspace{0.1cm} $\scriptstyle (\forall i )$ & \scalebox{1.2}{$ 2\bar{x}$} \hspace{0.1cm} $\scriptstyle (\forall i )$ \\
\hline
\scalebox{1.2}{$k$}&  \scalebox{1.2}{$2$}   & \scalebox{1.2}{$4 $} & \scalebox{1.2}{$ 1$}  & \scalebox{1.2}{$ 2$}\\
\hline
\end{tabularx}
\caption{Estimation variances associated to the $\mathbf{S}-$matrix, for three multiplexing-schemes. The variance expressions on pixel $i$ are valid for large number of pixels $N \gg 1$ (exact expressions in the proofs). $\mathbf{A}-$matrix: Global multiplexing matrix. $\mathbf{S_2}$: complementary matrix such that $\mathbf{S} + \mathbf{S_2} = \mathbf{J}_N$. Last column: balanced-detection strategy (section \ref{eqn:modelBalancedSubtract}). 
$\bar{x}$: object intensity average. $N$: object size. }
\label{tab:SumUp_Smat}
\end{table}

{\renewcommand{\arraystretch}{2}
\begin{table} [H]
\centering
\begin{tabularx}{\linewidth}{|p{1.8cm}|p{4.4cm}|p{4.4cm}|X|X|}
\hline
 &   One-step multiplexing & Two-step multiplexing  & \multicolumn{2}{|c|}{Dual-detection}  \\
\hline
$\mathbf{A}-$matrix &  $\mathbf{H_1}$ & $\mathbf{H_1} \otimes \mathbf{H_1}$  & $ \left[ \mathbf{H_1} \hspace{0.1cm} \mathbf{H_2} \right]^T$ & $\mathbf{H_1}-\mathbf{H_2}$ \\ 
\hline
\scalebox{1.2}{$V(\hat{x}_i)$} &  \scalebox{1.2}{$2 \bar{x}$} \hspace{0.1cm} $\scriptstyle (\forall i \neq 1 )$  & \scalebox{1.2}{$4 \bar{x}$} \hspace{0.1cm} $\scriptstyle (\forall i \neq n_1)$ & \multicolumn{2}{|c|}{\scalebox{1.2}{$ \bar{x}$} \hspace{0.1cm}  $\scriptstyle ( \forall i )$} \\
&   $ \textcolor{gray}{ (N-2) \bar{x}} $ \hspace{0.1cm} $\textcolor{gray}{\scriptstyle (i = 1)}$ & $\textcolor{gray}{ c' , d'} $ $\textcolor{gray}{ \scriptstyle (i = n_1) }$ & \multicolumn{2}{|c|}{} \\
\hline
\scalebox{1.2}{$k$}&  \scalebox{1.2}{$2$}   & \scalebox{1.2}{$4 $} & \multicolumn{2}{|c|}{\scalebox{1.2}{$ 1$}} \\
\hline
\end{tabularx}
\caption{Estimation variances associated to the positive-Hadamard matrix $\mathbf{H_1}$, for three multiplexing-schemes. The variance expressions on pixel $i$ are valid for large number of pixels $N \gg 1$ (exact expressions in the proofs). The results in black are valid on most pixels, while \textcolor{gray}{the results in  gray} are valid only on specific pixels ($n_1$: pixels on the first row and column of the image, $c',d'$: values given in equation \eqref{eqn:Vh1kronh1_result_Nlarge}. $\mathbf{A}-$matrix: Global multiplexing matrix. 
$\mathbf{H_2}$: complementary matrix such that $\mathbf{H_1} + \mathbf{H_2} = \mathbf{J}_N$. Last column: balanced-detection strategy (section \ref{eqn:modelBalancedSubtract}). 
$\bar{x}$: object intensity average. $N$: object size.}
\label{tab:SumUp_Hada}
\end{table}

\noindent Therefore, for a given one-step multiplexing scheme, positive-Hadamard multiplexing with the $\mathbf{S}-$ matrix or the $\mathbf{H_1}-$matrix lead to the same estimation variance on most object pixels. For both cases, two-step multiplexing leads to an estimation variance twice as large as one-step multiplexing. Implementing a dual-detection strategy for one-step multiplexing divides the estimation variance by 2, except when the balanced-detection strategy is employed for the $\mathbf{S}-$matrix. For $\mathbf{H_1}$, the balanced detection strategy  - defined in equation \ref{eqn:modelBalancedSubtract} - leads to approximately the same variance than when considering the full measurements (see section \ref{sec:JustificationBalancedDetection}). 


\subsection{Proofs of the variance results ($\mathbf{S}-$matrix)}
\label{sec:Multiplexing with the S-matrix}

Here we prove the results of Table \ref{tab:SumUp_Smat}. To do so, we simply insert matrices based on $\mathbf{S}-$matrix into the variance expressions of Fig.~\ref{fig:dessin_Modalities_General}.

\subsubsection*{One-step multiplexing}
\label{sec:Smat_Onestep multiplex}
Replacing $\mathbf{A}$ by $\mathbf{S}$ in equation  \eqref{eqn:OneStep_GeneralFormula} and using equations \eqref{eqn:S.S} and \eqref{eqn:Smatrixpty}, the estimation variance obtained for one-step multiplexing reads: 
\begin{align*}
\mathbf{V}_{S} (\hat{\mathbf{x}}) & = (\mathbf{S}^{-1} \odot \mathbf{S}^{-1})\mathbf{S}\mathbf{x} \\
& = \frac{4}{(N+1)^2} \mathbf{J_N} \mathbf{S}\mathbf{x}
\end{align*}
i.e. 
\begin{equation}
\boxed{\mathbf{V}_{S} (\hat{\mathbf{x}}) = \frac{2N}{N+1} \bar{x} \mathbf{1}_N}
\end{equation}
Therefore, if $N \gg 1$, the variance reads: 
\begin{equation}
\boxed{\mathbf{V}_{S} (\hat{\mathbf{x}}) \approx 2 \bar{x} \mathbf{1}_N}
\end{equation}
or, on every pixel $i$
\begin{equation}
V_S (\hat{x}_i)\approx 2 \bar{x}  \hspace{0.2cm} \forall i 
\end{equation}

\subsubsection*{Two-step multiplexing}
\label{sec:Smat_Twostep multiplex}
If the two multiplexing steps are based on S-matrices of size $\sqrt{N}$, then $\mathbf{P}^T = \mathbf{P} = \mathbf{U} = \mathbf{S}$. Replacing in \eqref{eqn:Kron_GeneralFormula} and using \eqref{eqn:Smatrixpty} and \eqref{eqn:S.S} lead to:
\begin{align*}
\mathbf{V}_{S \otimes S} (\hat{\mathbf{x}}) & = 
( (\mathbf{S}^{-1}  \odot  \mathbf{S}^{-1}) \mathbf{S} )  \otimes (\mathbf{S}^{-1}  \odot  \mathbf{S}^{-1}) \mathbf{S} ) )\mathbf{x} \\
& = \frac{16}{(\sqrt{N}+1)^4} (\mathbf{J}_{\sqrt{N}} \mathbf{S} \otimes \mathbf{J}_{\sqrt{N}} \mathbf{S}) \mathbf{x} \\
& = \frac{16}{(\sqrt{N}+1)^4} \frac{(\sqrt{N}+1)^2}{4} (\mathbf{J}_{\sqrt{N}} \otimes \mathbf{J}_{\sqrt{N}})\mathbf{x} \\
& = \frac{4}{(\sqrt{N}+1)^2}\mathbf{J}_N \mathbf{x} 
\end{align*}
i.e.
\begin{equation}
\boxed{\mathbf{V}_{S \otimes S} (\hat{\mathbf{x}}) = \frac{4N}{(\sqrt{N}+1)^2} \bar{x} \mathbf{1}_N}
\end{equation}
Therefore, if $N \gg 1$, the variance reads: 
\begin{equation}
\boxed{\mathbf{V}_{S \otimes S} (\hat{\mathbf{x}}) \approx 4 \bar{x} \mathbf{1}_N}
\end{equation}
or, on every pixel $i$
\begin{equation}
V_{S \otimes S} (\hat{x}_i) \approx 4 \bar{x} \hspace{0.2cm} \forall i  
\end{equation}
(We surmise that this result can be extended to $p-$dimensions (for $N \gg 1$): $\mathbf{V}_{S ^{\otimes p} } (\hat{\mathbf{x}})  \approx 2^p \bar{x} \mathbf{1}_{N^p}$).

\subsubsection*{Dual-detection}
\label{sec:Smat_Balanced}
In a dual detection scheme with a $\mathbf{S}-$matrix, one can define the two complementary matrices as: 
\begin{equation}
\left\{
    \begin{array}{ll}
        \mathbf{A_1} = \mathbf{S} \\
        \mathbf{A_2} = \mathbf{J}_N - \mathbf{S} \\
            \end{array}
\right.
\label{eqn:ModelBalancedSmat}
\end{equation}
We remind that estimation variance given in equation \eqref{eqn:VarianceGeneralBalanced_Block} is: 
\begin{align*}
\mathbf{V}(\hat{\mathbf{x}}) = (\mathbf{A}^{+} \odot \mathbf{A}^{+}) \mathbf{A} \mathbf{x}
\end{align*}
with
\begin{align*}
\mathbf{A} =  \begin{bmatrix}
\mathbf{A_1} \hspace{0.1cm}  \mathbf{A_2}
\end{bmatrix} ^T
\end{align*}
and 
\begin{align*}
\mathbf{A}^{+} =  (\mathbf{A}^{T} \mathbf{A})^{-1} \mathbf{A}^{T} = \left[ (\mathbf{A}^{T} \mathbf{A})^{-1} \mathbf{A_1}  \hspace{0.3cm}  (\mathbf{A}^{T} \mathbf{A})^{-1} \mathbf{A_2} \right]
\end{align*}
We have:
\begin{align*}
\mathbf{A}^T \mathbf{A} & = \mathbf{A_1}^T \mathbf{A_1} + \mathbf{A_2}^T \mathbf{A_2} \\
& = \mathbf{S}^T \mathbf{S} + \left( \mathbf{J}_N - \mathbf{S} \right)^T\left(\mathbf{J}_N -\mathbf{S}\right) \\
& =  2 \mathbf{S}\mathbf{S}  +  \mathbf{J}_N \mathbf{J}_N  - 2 \mathbf{J}_N \mathbf{S} \\
& = 2 \frac{N+1}{4} (\mathbf{I}_N +\mathbf{J}_N ) + N \mathbf{J}_N  - 2 \frac{N+1}{2} \mathbf{J}_N  \\
& = \frac{N+1}{2} \mathbf{I}_N  + \frac{N-1}{2} \mathbf{J}_N 
\end{align*}
Using equation \eqref{eqn:Sherman_Morrison} leads to:
\begin{align*}
\left( \mathbf{A}^T \mathbf{A} \right)^{-1} 
& = \frac{2}{N+1} \mathbf{I} - \frac{\left( \frac{2}{N+1} \right)^2 \frac{N-1}{2}}{1+\frac{2}{N+1}\frac{N-1}{2}N} \mathbf{J}_N  \\
& = \frac{2}{N+1} \mathbf{I}_N  - 2\frac{N-1}{N+1}\frac{1}{N^2+1} \mathbf{J}_N 
\end{align*}
Hence:
\begin{align*}
\left( \mathbf{A}^T \mathbf{A} \right)^{-1} \mathbf{A_1}
& = \frac{2}{N+1} \mathbf{S} - 2\frac{N-1}{N+1}\frac{1}{N^2+1}\frac{N+1}{2} \mathbf{J} \\
& = \frac{2}{N+1} \mathbf{S} - \frac{N-1}{N^2+1} \mathbf{J}_N 
\end{align*}
and
\begin{align*}
\left( \mathbf{A}^T \mathbf{A} \right)^{-1} \mathbf{A_2}
& = \frac{2}{N+1} \mathbf{J}_N - 2\frac{N-1}{N+1}\frac{N}{N^2+1} \mathbf{J}_N - \frac{2}{N+1} \mathbf{S} + \frac{N-1}{N^2+1} \mathbf{J}_N \\
& = \frac{N+1}{N^2+1}\mathbf{J}_N - \frac{2}{N+1} \mathbf{S}
\end{align*}
This leads to:
\begin{align}
\mathbf{A}^+  \odot \mathbf{A}^+
& = \left[ \left( \mathbf{A}^T \mathbf{C} \right)^{-1} \mathbf{A_1} \odot \left( \mathbf{A}^T \mathbf{A} \right)^{-1} \mathbf{A_1}
\ \hspace{0.5cm} \ 
\left( \mathbf{A}^T \mathbf{A} \right)^{-1} \mathbf{A_2} \odot \left( \mathbf{A}^T \mathbf{A} \right)^{-1} \mathbf{A_2} \right] \\
& = \left[
\alpha_1 \mathbf{S} + \beta_1 \mathbf{J}_N
\hspace{0.5cm}
\alpha_2 \mathbf{S} + \beta_2 \mathbf{J}_N
\right] 
\end{align}
with
\begin{align}
\left\{
    \begin{array}{ll}
\alpha_1 & = \frac{8}{(1+N)^2(1+N^2)} \\
\alpha_2 & = \frac{-8N}{(1+N)^2(1+N^2)} \\
\beta_1 & = \frac{(N-1)^2}{(1+N^2)^2} \\
\beta_2 & = \frac{(N+1)^2}{(1+N^2)^2}
            \end{array}
\right.
\end{align}
and thus (using equation \eqref{eqn:Smatrixpty}):
\begin{align*}
\left( \mathbf{A}^+  \odot \mathbf{A}^+ \right) \mathbf{A}
& = \alpha_1 \mathbf{S}\mathbf{S} + \beta_1 \mathbf{J}_N \mathbf{S} + \alpha_2 \mathbf{J}_N \mathbf{S} + \beta_2 \mathbf{J}_N\mathbf{J}_N - \alpha_2 \mathbf{S}\mathbf{S} - \beta_2 \mathbf{J}_N\mathbf{S} \\
& = \frac{N+1}{4} (\alpha_1-\alpha_2) \mathbf{I}_N +  \frac{N+1}{4} (\alpha_1+\alpha_2) \mathbf{J}_N + N\beta_2 \mathbf{J}_N + \frac{N+1}{2} (\beta_1-\beta_2) \mathbf{J}_N \\
& = \frac{2}{N^2+1} \mathbf{I}_N + \left( \frac{2(1-N)}{(N+1)(N^2 + 1)} + \frac{N(N+1)(N-1)}{(N^2 + 1)^2} \right) \mathbf{J}_N
\end{align*}
Finally, we can derive the variance: $\mathbf{V}_{Sd} (\hat{\mathbf{x}}) =  \left( \mathbf{A}^+  \odot \mathbf{A}^+ \right) \mathbf{A}\mathbf{x}$
\begin{equation}
\boxed{\mathbf{V}_{Sd} (\hat{\mathbf{x}}) = \frac{2}{N^2+1} \mathbf{x} + \left( \frac{2N(1-N)}{(N+1)(N^2 + 1)} + \frac{N^2(N+1)(N-1)}{(N^2 + 1)^2} \right) \bar{x} \mathbf{1}_N}
\end{equation}
Therefore, if  $N \gg 1$, the variance reads: 
\begin{equation}
\boxed{\mathbf{V}_{Sd} (\hat{\mathbf{x}}) \approx \bar{x} \mathbf{1}_N } \\
\label{eqn:Variance_Smat_DualDet}
\end{equation} \\

\noindent \underline{Note on the balanced detection strategy:} \\
As described in section~\ref{sec:noteBalancedDet}, one could as well subtract the two measurements and thus use the equivalent matrix: 
\begin{align*}
\mathbf{A} = \mathbf{A_1} - \mathbf{A_2}  = 2\mathbf{S} - \mathbf{J}_N = \frac{N+1}{2} \mathbf{S}^{-1} 
\end{align*}
Replacing in equation \eqref{eqn:Balanced_GeneralVarReal} leads to:  
\begin{align*}
\mathbf{V}_{Sb} (\hat{\mathbf{x}}) & = 
N \bar{x} \frac{4}{(N+1)^2}(\mathbf{S} \odot \mathbf{S}) \mathbf{1}_N = N \bar{x} \frac{4}{(N+1)^2} \frac{N+1}{2} \mathbf{1}_N 
\end{align*}
\begin{equation}
\mathbf{V}_{Sb} (\hat{\mathbf{x}})
= 2 \bar{x} \frac{N}{N+1} \mathbf{1}_N \\
\end{equation}
Therefore, if $N \gg 1$, the variance reads: 
\begin{equation}
\boxed{\mathbf{V}_{Sb} (\hat{\mathbf{x}}) \approx 2 \bar{x} \mathbf{1}_N}
\end{equation}
or, on every pixel $i$
\begin{equation}
V_{Sb} (\hat{x}_i) \approx 2 \bar{x} \hspace{0.2cm} \forall i  
\end{equation}
The obtained variance with is twice higher than the variance obtained (equation \eqref{eqn:Variance_Smat_DualDet}) with the dual detection strategy based on $\mathbf{A} = \left[
\mathbf{A_1} \hspace{0.1cm}  \mathbf{A_2}
\right]^T$. Therefore, when performing dual-detection with the $\mathbf{S}-$matrix, it is important \textit{not} to subtract the two measurements.

\subsection{Proofs of the variance results ($\mathbf{H_1}$-matrix)}
\label{sec:HadaPositive}
Here we prove the results of Table \ref{tab:SumUp_Hada}. To do so, we simply insert matrices based on the $\mathbf{H_1}-$ matrix into the expressions of Fig.~\ref{fig:dessin_Modalities_General}.

\subsubsection*{One-step multiplexing}
\label{sec:H1_Onestep multiplex}
Replacing $\mathbf{A}$ by $\mathbf{H_1}$ in equation  \eqref{eqn:OneStep_GeneralFormula}
and using equations \eqref{eqn:def_H1}, \eqref{eqn:H1_invert} and \eqref{eqn:JnH}, lead to the estimation variance obtained for one-step multiplexing: 
\begin{align*}
\mathbf{V}_{H1} (\hat{\mathbf{x}}) & = (\mathbf{H_1}^{-1} \odot \mathbf{H_1}^{-1})\mathbf{H_1}\mathbf{x} \\
& = (\frac{4}{N^2} \mathbf{J_N} + \frac{N-4}{N}\mathbf{e_1}\mathbf{e_1}^T )\mathbf{H_1}\mathbf{x} \\
& = (\frac{2}{N^2} \mathbf{J_N} + \frac{N-4}{2N}\mathbf{e_1}\mathbf{e_1}^T )(\mathbf{H} + \mathbf{J}_N) \mathbf{x} \\
& = (\frac{2}{N} \mathbf{1e_1}^T + \frac{N-4}{N} \mathbf{e_1 1}^T + \frac{2}{N} \mathbf{J}_N ) \mathbf{x}
\end{align*}
i.e. 
\begin{equation}
\boxed{\mathbf{V}_{H1} (\hat{\mathbf{x}}) = (2 \bar{x}  + \frac{2}{N} x_1 ) \mathbf{1}_N  +   (N-4) \bar{x} \mathbf{e_1}}
\end{equation}
If $N \gg 1$ and if $N \gg 2x_1$ (which can often be arranged in practice), the variance reads: 
\begin{equation}
\boxed{\mathbf{V}_{H1} (\hat{\mathbf{x}}) \approx 2 \bar{x}  \mathbf{1}_N  +  (N-4) \bar{x} \mathbf{e_1}}
\label{eqn:Vh1_result}
\end{equation}
or, on pixel $i$
\begin{equation}
V_{H1} (\hat{x}_i) 
\approx 
\left\{
    \begin{array}{ll}
        2 \bar{x}  & \mbox{} \forall i \neq 1  \\ 
       (N-2)\bar{x} & \mbox{for } i = 1
    \end{array}
\right.
\end{equation}
On all pixels but one, the variance equals twice the object average, as with the S-multiplexing. On the first object pixel ($i=1$), the variance scales with $(N-2)\bar{x}$, thus the first object pixel may often be mis-estimated. The influence of the first object pixel intensity $x_1$ is due to the structure of the Hadamard matrix with its first row and column with only ones. A similar derivation can be found in \cite{Studer2012}. 

\subsubsection*{Two-step multiplexing}
\label{sec:H1_Twostep multiplex}
If the two multiplexing steps are based on positive Hadamard-matrices of size $\sqrt{N}$, then $\mathbf{U} = \mathbf{P}^T = \mathbf{P} = \mathbf{H_1}$. Replacing in \eqref{eqn:Kron_GeneralFormula} and using the above result leads to:
\begin{align*}
\mathbf{V}_{H1 \otimes H1} (\hat{\mathbf{x}}) & = (((\mathbf{H_1}^{-1} \odot \mathbf{H_1}^{-1})\mathbf{H_1} ) \otimes(\mathbf{H_1}^{-1} \odot \mathbf{H_1}^{-1})\mathbf{H_1} ) \mathbf{x} \\
& = \frac{1}{N} ((2 \mathbf{1e_1}^T + (\sqrt{N} -4) \mathbf{e_1 1}^T + 2 \mathbf{J}_{\sqrt{N}}) \otimes (2\mathbf{1e_1}^T + (\sqrt{N} -4) \mathbf{e_1 1}^T + 2 \mathbf{J}_{\sqrt{N}})) \mathbf{x} \\
& = 4 \bar{x} \mathbf{1}_N + \frac{1}{N} (4 \mathbf{1e_1}^T \otimes  \mathbf{1e_1}^T + (\sqrt{N} -4)^2 \mathbf{e_1 1}^T \otimes \mathbf{e_1 1}^T  + 4 ( \mathbf{1e_1}^T \otimes \mathbf{J}_{\sqrt{N}} +  \mathbf{J}_{\sqrt{N}} \otimes \mathbf{1e_1}^T) + \\
& 2(\sqrt{N} - 4) ( \mathbf{1e_1}^T \otimes \mathbf{e_1 1}^T + \mathbf{e_1 1}^T \otimes\mathbf{1e_1}^T + \mathbf{e_1 1}^T \otimes \mathbf{J}_{\sqrt{N}} + \mathbf{J}_{\sqrt{N}} \otimes  \mathbf{e_1 1}^T ) \mathbf{x}
\end{align*}
As expected, it results in a constant term in $4\bar{x}$ and in many 'special pixels' given by the Kronecker products of the different elements. 
To isolate these special pixels it is relevant to treat the object as a 2-D object (which is the case in this two-step multiplexing scheme), as illustrated in Fig.~\ref{fig:Comparison_matrices_SI_2D}. We consider the matrix $\mathbf{X}$ with elements $X_{ij}$ where $\mathbf{x} = vec(\mathbf{X})$. 
Then, the variance reshaped in 2-D reads: 
\begin{equation}
\boxed{\mathbf{V}_{H1 \otimes H1} (\hat{\mathbf{X}}) = 4\bar{X} + \frac{4}{N}(X_{11}+\sum X_{1j}+ \sum X_{i1}) vec(\mathbf{J}_{\sqrt{N}})+  vec 
\begin{pmatrix}
c & d & ... & d\\
e & 0 & ... & 0 \\
. & 0 & ... & 0 \\
e & 0 & ... & 0 
\end{pmatrix}}
\label{eqn:Vh1kronh1_result}
\end{equation}
with
\begin{align*}
\left\{
\begin{array}{ll}
c&=(\sqrt{N} -4)(\sqrt{N}\bar{X}+ \frac{2}{N}(\sum X_{1j}+ \sum X_{i1})\\
d&=2(\sqrt{N} -4)(\bar{X}+ \frac{1}{N} \sum X_{i1})\\
e&= 2(\sqrt{N} -4)(\bar{X}+ \frac{1}{N} \sum X_{1j})
\end{array}
\right.
\end{align*}
where $\sum X_{i1}$ is the sum of all the elements of the first column of $\mathbf{X}$ and $\sum X_{1j}$ is the sum of all the elements of its first row. \\
\noindent When $N \gg 1$ (or more precisely, when $\sum X_{ij} \gg \sum X_{1j}$ and $\sum X_{ij} \gg \sum X_{i1}$, i.e. the sum of all the object elements is much larger than the sum of the elements of its first row or column), the variance simplifies to: 
\begin{equation}
\boxed{\mathbf{V}_{H1 \otimes H1} (\hat{\mathbf{X}}) \approx vec 
\begin{pmatrix}
c' & d' & ... & d'\\
d' & f' & ... & f' \\
. & f' & ... & f' \\
d' & f' & ... & f' 
\end{pmatrix}}
\label{eqn:Vh1kronh1_result_Nlarge}
\end{equation}
with
\begin{align*}
\left\{
\begin{array}{ll}
c'&=  4\bar{X} + \sqrt{N} (\sqrt{N} - 4)\bar{X}\\
d'&= 4 \bar{X} + 2(\sqrt{N} -4) \bar{X} \\
f'&= 4 \bar{X}
\end{array}
\right.
\end{align*}
In 2-D, under the above assumptions, the estimation variance is equal to 4 times the object average ($4\bar{X}$), on most pixels, except on its $1^{st}$ line and column. There, the $4\bar{X}$ value is supplemented by some constants ($c'$,$d'$,$e'$) that depend on the object average $\bar{X}$. This specific structure is illustrated in Fig.~\ref{fig:Comparison_matrices_SI_2D}.

\subsubsection*{Dual-detection}
\label{sec:H1_Balanced}
In a dual detection scheme with $\mathbf{H_1}$, one can define the two complementary matrices as: 
\begin{equation}
\left\{
    \begin{array}{ll}
        \mathbf{A_1} = \mathbf{H_1}  = \frac{1}{2}(\mathbf{J}_N + \mathbf{H}) \\
        \mathbf{A_2} = \mathbf{H_2} = \frac{1}{2}(\mathbf{J}_N - \mathbf{H}) \\
            \end{array}
\right.
\label{eqn:ModelBalancedHadamard}
\end{equation}
When considering the complete measurement vector and therefore the equivalent matrix $\mathbf{A} = \left[
\mathbf{H_1} \hspace{0.1cm}  \mathbf{H_2}
\right]^T$, and adapting the general calculation of section~\ref{sec:JustificationBalancedDetection} to the Hadamard matrix, the variance reads: 
\begin{equation}
\boxed{ \mathbf{V}_{Hd}(\hat{\mathbf{x}}) = \bar{x} (1 - \frac{N+3}{(N+1)^2}) \mathbf{1}_N + \frac{2}{N(N+1)} \mathbf{x} }
\end{equation}
Then, if $N \gg 1$: 
\begin{equation}
\boxed{ \mathbf{V}_{Hd}(\hat{\mathbf{x}}) \approx \bar{x} \mathbf{1}_N }
\end{equation}

\noindent \underline{Note for the balanced detection strategy:}\\
As described in section~\ref{sec:noteBalancedDet}, one could as well subtract the two measurements and thus use the equivalent matrix:
\begin{equation}
\mathbf{A} = \mathbf{H_1} - \mathbf{H_2} = \mathbf{H}
\end{equation}
The resulting equivalent matrix is therefore equal to the Hadamard matrix defined in equation~\eqref{eqn:Hadaproperty}.
Using equation~\eqref{eqn:Cov_Balanced_General} it is straighforward to show that the covariance matrix reads: 
\begin{equation}
\mathbf{\Gamma}_{H1b} = \bar{x} \mathbf{I}_N
\label{eqn:Cov_balanced}
\end{equation}
And therefore, the estimation variance reads:
\begin{equation}
\boxed{\mathbf{V}_{H1b} (\hat{\mathbf{x}}) = \bar{x} \mathbf{1}_N}
\label{eqn:Var_balanced_Hada}
\end{equation}
Or, on every pixel $i$:
\begin{equation}
V_{H1b} (\hat{x}_i) = \bar{x} \hspace{0.2cm}
\end{equation}

\noindent Hence, if $N \gg 1$, simply subtracting the measurements lead to a variance $\mathbf{V}_{H1b}$ approximately equal to the variance $\mathbf{V}_{H1d}$ obtained when considering the full measurement vector (equation \eqref{eqn:modelBalancedBlock}):
\begin{equation}
\boxed{\mathbf{V}_{H1b} (\hat{\mathbf{x}}) \approx \mathbf{V}_{H1d} (\hat{\mathbf{x}}) \approx \bar{x} \mathbf{1}_N}
\end{equation}
In both cases, the dual detection scheme with Hadamard matrices divides the variance by 2 as compared to one-step H1-multiplexing. 

\section{Positive-Cosine multiplexing}
\label{sec:Fourier}

Multiplexing based on positive-cosines (or positive-sines) modulation can be implemented in many different manners (e.g. \cite{Futia2011,Zhang2015,Hang2017,Zhang2017,
Meng2020, Moshtaghpour2018}). In an analogue way to positive-Hadamard-based multiplexing, an object intensity can be modulated with positive-cosine patterns, such as in Fig.~\ref{fig:FTIR_Interferometer}(a). Other cases of cosine-based multiplexing can be found in interferometric measurements (e.g. typical Michelson-interferometer of Fig.~\ref{fig:FTIR_Interferometer}(b)). 
In both cases, the measurement, in its discrete form, can be related to the object $\mathbf{x}$ by a general multiplexing matrix $\mathbf{C_1}$:
\begin{equation}
\mathbf{b} \sim Poisson (\mathbf{C_1} \mathbf{x})
\end{equation}

\begin{figure} [H]
\centering
\textbf{\includegraphics[scale=0.65]{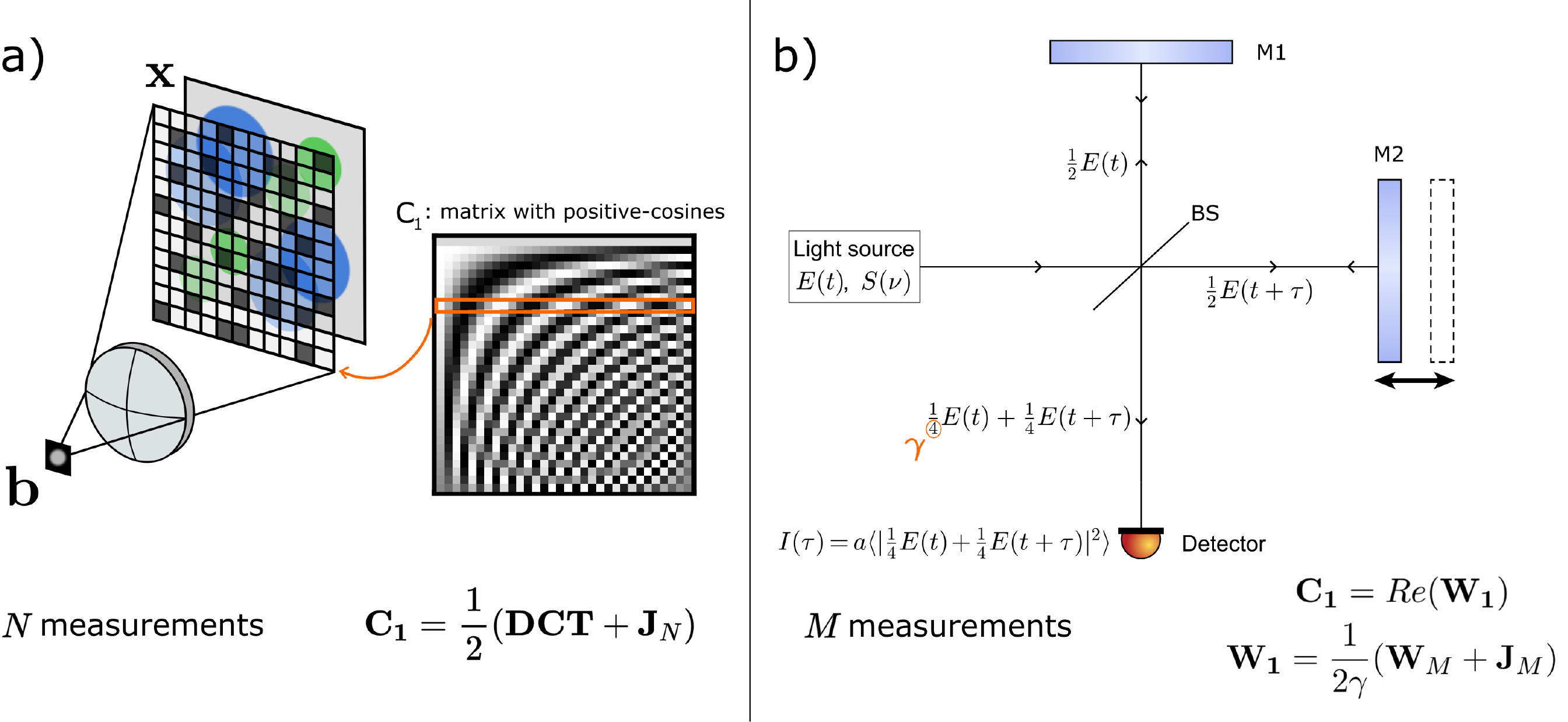} }
\caption{(a) Example of a positive-cosine modulation of an intensity object $\mathbf{x}$. (b) Scheme of a typical Michelson-interferometer. The detected intensity undergoes a modulation resulting from the interference of the non-monochromatic electric field $E(t)$ at several delays $\tau$. $I(\tau)$: detected intensity, $S(\nu)$: power spectrum of $E(t)$, BS: beam splitter, M1: fixed mirror, M2: moving mirror inducing time delay $\tau$ (see section \ref{sec:W1_PhysicalPicture} for more details).}
\label{fig:FTIR_Interferometer}
\end{figure}

\subsubsection*{Case 1: Fig.~\ref{fig:FTIR_Interferometer}(a)}
\noindent Multiplexing based on intensity modulation is often implemented on systems similar Fig.~\ref{fig:FTIR_Interferometer}(a).
In an analogue way to positive-Hadamard-based multiplexing (section~\ref{sec: Hadamard-based multiplexing}, an object intensity of size $N$ is modulated with $N$ positive-cosine patterns. The patterns generally derive from the discrete cosine transform (DCT) matrix, but the latter has different possible definitions \cite{Strang1999}.
The most familiar definitions of DCT are equivalent - to some normalisation factor - to the discrete Fourier transform of real numbers with even symmetry \cite{Strang1999}. But the MSE results may depend which DCT definition is chosen. Here, we choose the multiplexing matrix $\mathbf{C_1}$ such that its coefficients are comprised between 0 and 1:
\begin{equation}
\mathbf{C_1} = \frac{1}{2} (\mathbf{DCT} + \mathbf{J}_N)
\label{eqn:C1_DCT_def}
\end{equation}
where $\mathbf{DCT}$ is the discrete-Cosine transform matrix with
coefficients comprised between -1 and +1. \\

\subsubsection*{Case 2: Fig.~\ref{fig:FTIR_Interferometer}(b)}
\noindent Other types of systems, such as interferometric systems \cite{fellgett1951theory, Fuhrmann2004} or \cite{Futia2011, Scotte2019}, also perform cosine-based multiplexing. For example, in the typical Michelson-interferometer of Fig.~\ref{fig:FTIR_Interferometer}(b), the detected intensity $I(\tau)$ undergoes a modulation, which is related by some positive-cosine transform to the the field power spectrum $S(\nu)$ (section \ref{sec:W1_PhysicalPicture}). In such systems, the matrix $\mathbf{C_1}$ is based on the real part of the Fourier matrix, and the number of measurements is assumed to be $M \geq N$. In addition, there exist many different interferometers layouts and system specificities that lead to different equivalent multiplexing matrices $\mathbf{C_1}$ \cite{Treffers1977, Fuhrmann2004}. 
For such systems, we define generic positive-cosine multiplexing matrix $\mathbf{C_1}$ related to the real part of a matrix $\mathbf{W_1}$: 
\begin{equation}
\mathbf{W_1} = \frac{1}{2\gamma} (\mathbf{W}_M + \mathbf{J}_M)
\label{eqn:W1_def}
\end{equation}
where $\mathbf{W}_{M} \in \mathbb{C}_{M^2}$ is the discrete inverse Fourier transform matrix defined in equation \eqref{eqn:Wdef} and $\gamma$ is a real positive constant that accounts for some system specificities (for example, for the ideal Michelson interferometer of Fig.~\ref{fig:FTIR_Interferometer}(b), $\gamma = 4$ (section \ref{sec:W1_PhysicalPicture}).  Here we restrict ourselves to the case $M \approx 2N$. 


\subsection{Results}
The resulting estimation variances are shown in Table.~\ref{tab:SumUp_Fourier}. Note that the variances of the system of Case 1 and Case 2 cannot directly be compared since the number of measurements is not the same. 
For alternative definitions of the multiplexing matrix $\mathbf{C_1}$, the resulting variances can be derived using Table.~\ref{tab:PertubEpsilon} and Table.~\ref{tab:Four scenarios}. Overall, the variance associated with a general positive-cosine matrix is equal to a constant which depends on the exact definition of $\mathbf{C_1}$. \\ 

\noindent Importantly, we note that for Case 1, the variance is twice larger than for positive-Hadamard multiplexing with $\mathbf{H_1}$ (Table.~\ref{tab:SumUp_Hada}). This may be surprising but can be explained  because $\mathbf{H_1}$ and $\mathbf{C_1}$ - as defined in equation~\eqref{eqn:def_H1} and \eqref{eqn:C1_DCT_def}, respectively - do not have the same norm \footnote{largest singular value}. This means that, for an identical system, the energy transmitted by $\mathbf{H_1}$ is not the same as the energy transmitted by $\mathbf{C_1}$. A matrix $\mathbf{C_1}$ defined such that its norm is equal to the norm of $\mathbf{H_1}$ leads to the same variance as with $\mathbf{H_1}$, but would comprise some negative coefficients.\\

\noindent For Case 2, the proofs are provided in section \ref{sec:proofsFourier}. From these proofs, the variance for Case 1 is deduced, and is verified numerically in Fig.~S2 (Supp Methods).

{\renewcommand{\arraystretch}{2}
\begin{table} [H]
\centering
\begin{tabularx}{\linewidth}{|p{1.8cm}|p{4.4cm}|p{4.4cm}|X|X|}
\hline
 &   One-step multiplexing & Two-step multiplexing  & \multicolumn{2}{|c|}{Dual-detection}  \\
\hline
$\mathbf{A}-$matrix &  $\mathbf{C_1}$ & $\mathbf{C_1} \otimes \mathbf{C_1}$  & $\left[ \mathbf{C_1} \hspace{0.1cm} \mathbf{C_2} \right]^T$ & $\mathbf{C_1}-\mathbf{C_2}$ \\ 
\hline \hline
\multicolumn{5}{|c|}{General positive-cosine matrix $\mathbf{C_1}$}\\
\hline
\scalebox{1.2}{$V(\hat{x}_i)$} & \scalebox{1.2}{$k' \bar{x}$}  & \scalebox{1.2}{$k' \bar{x}$}  & \multicolumn{2}{|c|}{\scalebox{1.2}{$k'' \bar{x}$}}  \\
\hline \hline
\multicolumn{5}{|c|}{Case 1: $\mathbf{C_1}$ of size $N \times N$, with coefficients between 0 and 1. $\mathbf{C_1} = \frac{1}{2}(\mathbf{DCT} + \mathbf{J}_N)$}\\
\hline
\scalebox{1.2}{$V(\hat{x}_i)$} & \scalebox{1.2}{$4 \bar{x}$}  & \scalebox{1.2}{$16 \bar{x}$}  & \multicolumn{2}{|c|}{\scalebox{1.2}{$2 \bar{x}$}}  \\
\hline \hline
\multicolumn{5}{|c|}{Case 2: $\mathbf{C_1}$ of size $M \times M$, related to real part of $\mathbf{W_1}=\frac{1}{2\gamma} (\mathbf{W}_M + \mathbf{J}_M)$} \\
\hline
\scalebox{1.2}{$V(\hat{x}_i)$} & \scalebox{1.2}{$ 2\gamma \bar{x}$} \hspace{0.1cm} $\scriptstyle (\forall i \neq 1 ) $  & \scalebox{1.2}{$ 4\gamma^2 \bar{x}$}  \hspace{0.1cm} $\scriptstyle ( \forall i \neq  n_1 )$ & \multicolumn{2}{|c|}{\scalebox{1.2}{$ \gamma \bar{x}$} \hspace{0.1cm}  $\scriptstyle (\forall i \neq 1 )$} \\
 &   $ \textcolor{gray}{ 2\gamma (M-2) \bar{x}  }$ \hspace{0.1cm} $\textcolor{gray}{  \scriptstyle ( i = 1 )}$  & $\textcolor{gray}{c', d'}$\hspace{0.1cm} $ \textcolor{gray}{ \scriptstyle ( \forall i \neq  n_1 ) }$  & \multicolumn{2}{|c|}{$ \textcolor{gray}{ 2\gamma \bar{x} }$  $ \textcolor{gray}{ \scriptstyle( i = 1 )}$} \\
\hline
\end{tabularx}
\caption{Multiplexing-schemes and associated estimation variances in the three multiplexing modalities . The variance expressions on pixel $i$ are valid for large number of pixels $N \gg 1$ (exact expressions in the proofs). The results in black are valid on most pixels, while \textcolor{gray}{the results in  gray} are valid only on specific pixels. $k'$, $k''$ and $k'''$ are constants that depend on the exact definition of $\mathbf{C_1}$. $\gamma \in \mathbb{R}_{+}^*$ is a constant. $\mathbf{A}-$matrix: Global multiplexing matrix. $\mathbf{C_2}$ is the complementary matrix of $\mathbf{C_1}$. The last column corresponds to a balanced-detection strategy (section \ref{eqn:modelBalancedSubtract}). 
$\bar{x}$: object intensity average, $N$: object size, $M=2(N-1)$, $n_1$: specific pixels on the first row and column of the image,  $c',d'$ are given in equation \eqref{eqn:Vw1kronw1_result_Nlarge}.}
\label{tab:SumUp_Fourier}
\end{table}

\noindent \underline{Note:} For positive-cosine multiplexing, common alternative solutions to the balanced-detection scheme presented here exist (e.g. the four-step phase-shifting method), with the aim of removing the detected DC component \cite{Zhang2015, Meng2020}. Such strategies can improve the MSE (at the expense of a higher number of measurements), but are not considered in this work. 
 

\subsection{Proofs for Case 2 (Fig.~\ref{fig:FTIR_Interferometer}(b))}

\subsubsection*{Model: positive-multiplexing via interferometric measurements}
\label{sec:W1_PhysicalPicture}
First, we consider the physical model Michelson-based interferometric measurements. We assume that we seek to estimate a power spectrum $S(\nu)$ from intensity measurements $I(\tau)$ at different time delays $\tau$.
Here, the constant $\gamma$ may depend on the specific method of modulation and detection \cite{Treffers1977, Fuhrmann2004}, for instance on the interferometer design, on the beam-splitter, etc. 
To a first approximation (perfect beam-splitter, no apodization, etc), the measured intensity $I(\tau)$ by the above system reads: 
\begin{equation}
I(\tau) = \int_{-\infty}^{\infty} |\frac{1}{\sqrt{4\gamma}} E(t) + \frac{1}{\sqrt{4\gamma}} E(t+\tau)|^2 \, \mathrm{d}t
= \frac{1}{2\gamma} (I_0 + R(\tau))
\end{equation} 
where $I_0$ is the intensity of the initial field $E(t)$ :
\begin{equation}
I_0 = \int_{-\infty}^{\infty} |E(t)|^2 \, \mathrm{d}t
= \int_{-\infty}^{\infty} S(\nu) \, \mathrm{d}\nu
\label{eqn:Parseval}
\end{equation} 
and $R(\tau)$ is the field autocorrelation, i.e. the inverse Fourier-transform of the power spectrum $S(\nu)$:
\begin{equation}
R(\tau) = \int_{-\infty}^{\infty} E(t) E(t+\tau) \, \mathrm{d}t 
= \int_{-\infty}^{\infty} S(\nu)e^{2i\pi\nu\tau} \, \mathrm{d}\nu 
\end{equation} 
Therefore, the detected intensity becomes:
\begin{equation}
I(\tau) = \frac{1}{2\gamma} (I_0 + \int_{-\infty}^{\infty} e^{2i\pi\nu\tau} S(\nu) \, \mathrm{d}\nu  ) = \frac{1}{2\gamma} \int_{-\infty}^{\infty} (1 + e^{2i\pi\nu\tau}) S(\nu) \, \mathrm{d}\nu 
\label{eqn:FourierIntensityContinous}
\end{equation} 
The least-square estimation of the power spectrum is thus given by the real part of the Fourier-transform: 
\begin{equation}
\hat{S}(\nu) = Re (\int_{-\infty}^{\infty} (2\gamma I(\tau) - I_0) e^{-2i\pi\nu\tau} \, \mathrm{d}\tau )
\label{eqn:FourierEstimateContinous}
\end{equation} 
Since $2\gamma I(\tau) - I_0 \in \mathbb{R}$ and  $\hat{S}(\nu)$ is even, the physical power spectrum is thus obtained by selecting only the positive frequencies, and multiplying the intensity by 2 in order to fulfil \eqref{eqn:Parseval}. 
Note that then, the number of spanned delays $\tau$ should be at least twice higher than the number of spanned frequencies $\nu$.

\subsubsection*{Associated discrete model}
The above model can be written is a discrete form using the following variables: 
\begin{itemize}
\item $\mathbf{b} \in \mathbb{N}_{M}^+$: photon counts of the detected intensity $I(\tau_m)$ with $\tau_m = 0:M-1$ 
\item $\mathbf{W}_{M} \in \mathbb{C}_{M^2}$: discrete inverse Fourier transform matrix, defined in equation \eqref{eqn:Wdef}. 
\item $\mathbf{x} \in \mathbb{R}_N^+$: discrete power spectrum $S(\nu_n)$ with $\nu_n \in [0:N-1]$
\item  $\mathbf{x}^s \in \mathbb{R}_{M}^+$: symmetrized discrete power spectrum.
\end{itemize}
Therefore, we consider the following model: 
\begin{equation}
\mathbf{b} \sim Poisson (\mathbf{W_1} \mathbf{x}^s )
\end{equation}
where $\mathbf{W_1}$ is defined as in equation~\ref{eqn:W1_def} 
$\mathbf{b}$ is the measurement vector of size $M = 2(N-1)$. $\mathbf{x}_s \in \mathbb{R}_{M}^+$ is a temporary variable built from $\mathbf{x}$ and its even-symmetric: \\
\begin{equation}
\mathbf{x}^s = \frac{1}{2} [x_0,x_1,...,x_{N-1},x_{N-2},...x_1]^T. 
\label{eqn:defSymmetricObject}
\end{equation}
Then, even if $\mathbf{W_1}$ is a complex quantity, $\mathbf{W_1} \mathbf{x}_s $ is a real positive quantity \footnote{If $g$ is real and even, its Fourier-transform is real and even.} and so the measurement $\mathbf{b}$ can be defined ($\mathbf{b} \in \mathbb{N}^N$). 
Note that we define $\mathbf{x}^s$ so that $\mathbf{x}$ and $\mathbf{x}^s$ have the same energy: 
\begin{equation}
I_0 = \sum x_n = \sum x^s_n = N \bar{\mathbf{x}} = 2(N-1) \bar{\mathbf{x}}^s
\label{eqn:I0Fourier}
\end{equation}
The LS-estimate of $\mathbf{x}^s$ is: 
\begin{equation}
\hat{\mathbf{x}}^s = \mathbf{W_1}^{-1} \mathbf{b} = \frac{1}{2}[\hat{x}_0...\hat{x}_{N-1},\hat{x}_{N-2},...,\hat{x}_1]^T
\end{equation}
The final estimate, denoted $\hat{\mathbf{x}}_{Re}$ is obtained by taking the real part of $\hat{\mathbf{x}}^s$, selecting half of it and multiplying it by 2 (according to equation \eqref{eqn:defSymmetricObject}) \footnote{Since $\mathbf{b}$ is real, $Re(\hat{\mathbf{x}}^s)$ is even, therefore this is equivalent to averaging both parts of $Re(\hat{\mathbf{x}}^s)$.}
\begin{equation}
\hat{\mathbf{x}}_{Re} = Re \left( \hat{\mathbf{x}} \right) = 2 Re(\hat{\mathbf{x}}^s)_{[0:N-1]} = 
Re[\hat{x}_{0},...,\hat{x}_{N-1}]^T
\label{eqn:EstimateRealx}
\end{equation}


\subsubsection*{The discrete Fourier transform (DFT) matrix}
\label{sec:W1_Properties}
$\mathbf{W}_{M} \in \mathbb{C}_{M^2}$ is the inverse of the discrete Fourier transform matrix. Its $(k,m)-$element reads, for $k,m = ( 0, 1, ... , M-1)$: 
\begin{equation}
w_{M}^{km} = e^{\frac{2 i\pi k m}{M}} \\
\label{eqn:Wdef}
\end{equation}\\
\begin{equation*}
\mathbf{W}_{M} = \begin{pmatrix}
1 & 1 & 1 & \cdots & 1 \\
1 & w_{M} & w_{M}^{2} & \cdots & w_{M}^{M-1} \\
\vdots  & \vdots  & \vdots & \ddots & \vdots  \\
1 & w_{M}^{M-1} & w_{M}^{2(M-1)} & \cdots & w_{M}^{(M-1)^2} \\
\end{pmatrix}
\end{equation*}
It is a symmetric matrix ($\mathbf{W}_{M}^{T} = \mathbf{W}_M$) with properties similar to the Hadamard matrix $\mathbf{H}$:  
\begin{equation}
\mathbf{W}_{M}\mathbf{W}_M^* = M \mathbf{I}_M
\label{eqn:Wpty_In}
\end{equation}
\begin{equation}
\mathbf{W}_M \odot \mathbf{W}_M^* = \mathbf{J}_M
\label{eqn:W.W}
\end{equation}
\begin{equation}
\mathbf{J}_N \mathbf{W}_N = \mathbf{J}_N \mathbf{W}_N^* = N \mathbf{1}\mathbf{e_1}^T 
\label{eqn:JnW}
\end{equation}
\begin{equation}
\mathbf{W}_N \mathbf{J}_N  = \mathbf{W}_N^* \mathbf{J}_N  = N \mathbf{e_1} \mathbf{1}^T 
\label{eqn:WJn}
\end{equation}
$\mathbf{W_1}$ is the complex analogous of the positive Hadamard matrix $\mathbf{H_1}$. Using \eqref{eqn:Sherman_Morrison}, its inverse reads: 
\begin{equation}
\mathbf{W_1}^{-1} = 2 \gamma(\mathbf{W}_M + \mathbf{J}_M)^{-1} = \frac{2 \gamma }{M}(\mathbf{W}_M^* - \frac{M}{2} \mathbf{e_1}\mathbf{e_1}^T) 
\label{eqn:W1_invert}
\end{equation}


\noindent In addition, since $\mathbf{W}_M \in \mathbb{C}_{M^2}$, we need to consider the estimation variance in the complex case. 
Using the complex definition of the covariance ($\mathbf{\Gamma} = \langle \delta\hat{\mathbf{x}}\delta\hat{\mathbf{x}}^{H}\rangle$) and equation \eqref{eqn:diag_propertyCmplx}, the estimation variance reads: 
\begin{equation}
\mathbf{V}(\mathbf{\hat{x}}) = (\mathbf{A}^{-1} \odot (\mathbf{A}^{-1})^*)\mathbf{A}\mathbf{x}
\label{eqn:VarianceGeneralFormula_Cmplx}
\end{equation}
The variance of the estimate real part $\mathbf{\hat{x}}_{Re} = Re(\mathbf{\hat{x}})$ is:  
\begin{equation}
\mathbf{V}(\mathbf{\hat{x}}_{Re}) = \frac{1}{2} Re (diag \hspace{0.1cm}(\langle \delta\mathbf{\hat{x}}\delta\mathbf{\hat{x}}^{H} \rangle + \langle \delta\mathbf{\hat{x}}\delta\mathbf{\hat{x}}^T \rangle )) = \frac{1}{2} Re \left(  (\mathbf{A}^{-1} \odot (\mathbf{A}^{-1})^* + \mathbf{A}^{-1} \odot \mathbf{A}^{-1})\mathbf{A}\mathbf{x} \right)
\label{eqn:VarianceGeneralFormula_ComplxRealPart}
\end{equation}

\subsubsection*{One-step multiplexing}
\label{sec:proofsFourier}
\label{sec:W1_Onestep multiplexing}
To calculate the variance of the real part of $\hat{\mathbf{x}}$, we first calculate the variance of the real part of the symmetrized object $\hat{\mathbf{x}}^s$. It is obtained by inserting $\mathbf{W_1}^{-1}$ into \eqref{eqn:VarianceGeneralFormula_ComplxRealPart} : 
\begin{align*}
\mathbf{V}_{W1}(\hat{\mathbf{x}}_{Re}^s) & = \frac{2\gamma}{2M^2} Re \left( \left( (\mathbf{W}_M^* - \frac{M}{2} \mathbf{e_1}\mathbf{e_1}^T) \odot (\mathbf{W}_M - \frac{M}{2} \mathbf{e_1}\mathbf{e_1}^T) + (\mathbf{W}_M^* - \frac{M}{2} \mathbf{e_1}\mathbf{e_1}^T)^{\odot 2}  \right)
(\mathbf{W}_M + \mathbf{J}_M) \mathbf{x}^s \right)\\
& = \frac{2\gamma}{2M^2} Re \left( 
(\mathbf{J}_M + \mathbf{W}_M^{*} \odot \mathbf{W}_M^{*} + \frac{M(M-4)}{2} \mathbf{e_1}\mathbf{e_1}^T )(\mathbf{W}_M + \mathbf{J}_M) \mathbf{x}^s \right)\\
& = \frac{2\gamma}{2M^2} Re \left( 
( M \mathbf{J}_M + \mathbf{J}_M \mathbf{W}_M + \widetilde{\mathbf{W}}_M^{*} \mathbf{W}_M +  \widetilde{\mathbf{W}}_M^* \mathbf{J}_M +  M^2 (M-4) \mathbf{e_1}\mathbf{1}^T)\mathbf{x}^s \right)
\end{align*}
where we used the fact that  $\mathbf{e_1}\mathbf{e_1}^T\mathbf{W}_M = \mathbf{e_1}\mathbf{e_1}^T\mathbf{J}_M = \mathbf{e_1}\mathbf{1}_M^T$, and defined $\widetilde{\mathbf{W}}_M^* = \mathbf{W}_M^{*} \odot \mathbf{W}_M^{*}$.\\
Each matrix element of $\widetilde{\mathbf{W}}_M^* \mathbf{W}_M $ is: 
\begin{equation}
r_{km} = \sum_{h=0}^{M-1} (e^{\frac{2 i \pi}{M}(m-2k)})^h 
= 
\left\{
    \begin{array}{ll}
        M & \mbox{if } m-2k = 0 (M) \\
        0 & \mbox{otherwise.}
    \end{array}
\right.
\end{equation}
In other words, the matrix $\widetilde{\mathbf{W}}_M^* \mathbf{W}_M$ selects only the even pixels of $\mathbf{x}^s$ in a symmetric manner. Therefore, we can write
\begin{equation}
(\widetilde{\mathbf{W}}_M^* \mathbf{W}_M ) \mathbf{x}^s = M 
\begin{bmatrix}
\mathbf{v}\\
\mathbf{v}
\end{bmatrix}
\end{equation}
where 
\begin{align*}
\mathbf{v} = (x^s_0 , x^s_2 , ... , x^s_{M-2}, x^s_0 , x^s_2 , ... , x^s_{M-2})^T 
\end{align*}
In addition, using \eqref{eqn:WJn} and noting that all the terms are real quantities, leads to:
\begin{equation}
\mathbf{V}_{W1}(\hat{\mathbf{x}}_{Re}^s) =
\gamma
\bar{x}^s \left( \mathbf{1}_M +  \begin{bmatrix}
\mathbf{e_1}_{,N}\\
\mathbf{e_1}_{,N}
\end{bmatrix} + (M-4) \begin{bmatrix}
\mathbf{e_1}_{,N}\\
\mathbf{0}_{N}
\end{bmatrix} \right) + \frac{\gamma}{M} \left( x_0^s\mathbf{1}_M + \begin{bmatrix}
\mathbf{v}\\
\mathbf{v}
\end{bmatrix} \right)
\label{eqn:VarianceFourierReal_symmetric}
\end{equation}
Then, $\hat{\mathbf{x}}_{Re} $ is obtained from  $\hat{\mathbf{x}}_{Re}^s $ with equation \eqref{eqn:EstimateRealx}. Further applying $\bar{x}^s \approx \bar{x}/2$ (equation \eqref{eqn:I0Fourier}, if $N \gg 1$)) leads to an overall variance  of: 
\begin{equation}
\boxed{\mathbf{V}_{W1}(\hat{\mathbf{x}}_{Re})  = 2 \gamma
\bar{x} (\mathbf{1}_N + (M-3)
\mathbf{e_1}_{,N}) + \frac{2 \gamma}{M} (x_0 \mathbf{1}_N +
2 \mathbf{v} )}
\end{equation}
Therefore, if $M$ is large compared to the components of vector $\mathbf{v}$: 
\begin{equation}
\boxed{\mathbf{V}_{W1}(\hat{\mathbf{x}}_{Re})  \approx 2\gamma
\bar{x} (\mathbf{1}_N + (M-3)
\mathbf{e_1}_{,N}) }
\end{equation}
i.e. on pixel $i$:
\begin{equation}
V_{W1}(\hat{x}_{i, Re}) 
\approx 
\left\{
    \begin{array}{ll}
        2 \gamma
\bar{x} & \mbox{} \forall i \neq 0 \\
        2 \gamma (M-2)
\bar{x} & \mbox{for }  i = 0
    \end{array}
\right.
\end{equation}
Note that an error is also present in the imaginary part of the estimate $\hat{\mathbf{x}}^s$. A similar but simpler calculation shows that the variance of the complex estimate $\hat{\mathbf{x}}^s$ is twice higher that in the above expressions: the variance seems to spread equally in the real and imaginary part of the estimate.

\noindent 
Note also that if the total energy $I_0=\sum \mathbf{x}$ is known \textit{a priori}, the large variance of the first pixel can be attenuated: the estimation variance is the same as above, except for the term proportional to $(M-2)\bar{x}\mathbf{e_1}$. Indeed, in this case, the model reads: 
\begin{equation}
\mathbf{b} \sim Poisson (\frac{1}{2\gamma}(I_0 \mathbf{1}_M + \mathbf{W}_M \mathbf{x}^s) )  
\end{equation}
The estimate is: 
\begin{equation}
\hat{\mathbf{x}}^s = \mathbf{W}_M^{-1} (2\gamma \mathbf{b} - I_0 \mathbf{1}_{M})  
\end{equation}
Using the same kind of calculations as above, when if $M$ is large compared to the components of vector $\mathbf{v}$, the variance of the estimate real part reads: 
\begin{align*}
\mathbf{V}(\hat{\mathbf{x}}_{Re})  \approx 2 \gamma
\bar{x} \mathbf{1}_N 
\end{align*}

\subsubsection*{Two-step multiplexing}
\label{sec:W1_Twostep multiplexing}
In the case of two-step multiplexing with positive Fourier-matrices of size $\sqrt{N}$, the estimation variance is obtained by replacing $\mathbf{U}$ and  $\mathbf{P}$
by $\mathbf{W_1}$ in \eqref{eqn:Kron_GeneralFormula}. Using \eqref{eqn:VarianceGeneralFormula_ComplxRealPart}, the general formula for the variance reads:
\begin{equation}
\mathbf{V}_{W1 \otimes W1}(\hat{\mathbf{x}}_{Re}^s)  = \frac{1}{2} Re \left ( \hspace{0.1cm}((\mathbf{W_1}^{-1} \odot (\mathbf{W_1}^{-1})^*)\mathbf{W_1} )^{\otimes 2} +  
((\mathbf{W_1}^{-1} \odot \mathbf{W_1}^{-1})\mathbf{W_1} ) ^{\otimes 2} \hspace{0.1cm} \right) \\
\end{equation}
Inserting \eqref{eqn:W1_invert} and performing a similar calculation that in section~\ref{sec:H1_Twostep multiplex} with the 2D-object $\mathbf{X}$ defined by $\mathbf{x} = vec(\mathbf{X})$ leads to: 
\begin{equation}
\mathbf{V}_{W1 \otimes W1}(\hat{\mathbf{x}}_{Re})  = \frac{4 \gamma^2}{4} \left( 4\bar{X} + \frac{4}{M}(X_{00}+\sum X_{0j}+ \sum X_{i0}) vec(\mathbf{J}_{\sqrt{N}}) +
vec 
\begin{pmatrix}
c & d_1 & \cdots & d_{N-1}\\
e_1 & f_{11} & \cdots & f_{1(N-1)} \\
\vdots & f_{21} & \cdots & f_{2(N-1)} \\
e_{N-1} & f_{(N-1)1} & \cdots & f_{(N-1)(N-1)} 
\end{pmatrix} 
\right)
\label{eqn:Vw1kronh1_result}
\end{equation}
where 
\begin{align*}
\left\{
    \begin{array}{ll}
c&= 2(M - 4\sqrt{M} + 4) \bar{X} +  \frac{4}{M} (\sqrt{M}-3) (\sum X_{0j}+ \sum X_{i0}) +  \frac{4}{M} X_{00} \\
d_j& = 2(\sqrt{M} -4)\bar{X} + \frac{2}{M} ( (\sqrt{M}-4)\sum X_{i0} + 2(\sqrt{M}-2) \sum X_{0(2j)}^s ) + \frac{8}{M} X_{0(2j)}^s \\
e_i& = 2(\sqrt{M} -4)\bar{X} + \frac{2}{M} ( (\sqrt{M}-4)\sum X_{0j} + 2(\sqrt{M}-2) \sum X_{(2i)0}^s ) + \frac{8}{M} X_{(2i)0}^s \\
f_{ij} & =  \frac{8}{M} X_{(2i)(2j)}^s
\end{array}
\right.
\end{align*}
As expected, the variance results in a constant term in $\gamma^2 \bar{X}$ and in many 'special pixels' given by the Kronecker products of the different elements. 
$\sum X_{i0}$ is the sum of all the elements of the $0^{th}$ column of $\mathbf{X}$, $\sum X_{0j}$ is the sum of all the elements of its $0^{th}$ row,  $X_{(2i)(2j)}^s$ is an element of a even row and column of the symmetrised object $\mathbf{X}^s$. 
When $M \gg 1$, the variance simplifies to: 
\begin{equation}
\boxed{\mathbf{V}_{W1 \otimes W1} (\hat{\mathbf{x}}_{Re}) \approx  vec 
\begin{pmatrix}
c' & d' & ... & d'\\
d' & f' & ... & f' \\
. & f' & ... & f' \\
d' & f' & ... & f' 
\end{pmatrix}}
\label{eqn:Vw1kronw1_result_Nlarge}
\end{equation}
where 
\begin{align*}
\left\{
    \begin{array}{ll}
c'&=  4 \gamma^2 \bar{X} + 2 \gamma^2 (M - 4 \sqrt{M} + 4) \bar{X}\\
d'&= 4 \gamma^2 \bar{X} + 2 \gamma^2 (\sqrt{M} -4) \bar{X} \\
f'&= 4 \gamma^2 \bar{X}
\end{array}
\right.
\end{align*}
The 2-D image of the above variance is thus approximately equal to $\gamma^2 \bar{X}$ on most pixels, except on its $1^{st}$ row, $1^{st}$ column, and one some specific pixels which influence is negligible if $M \gg 1$ (see for example Fig.~\ref{fig:Comparison_matrices_SI_2D}).

\subsubsection*{Dual detection}
\label{sec:W1_Balanced}
In positive-cosine multiplexing, many different strategies may be implemented to remove the DC component, such as four-step-shifting detection \cite{Zhang2015, Meng2020}. Here, we restrict ourselves to the dual detection as defined in section~\ref{sec: Balanced detection}. In a dual detection scheme with $\mathbf{W_1}$, one can define the two complementary matrices as: 
\begin{equation}
\left\{
    \begin{array}{ll}
        \mathbf{A_1} = \mathbf{W_1}  = \frac{1}{2\gamma}(\mathbf{J}_M + \mathbf{W}_M) \\
        \mathbf{A_2} = \mathbf{W_2} = \frac{1}{2\gamma}(\mathbf{J}_M - \mathbf{W}_M) \\
            \end{array}
\right.
\label{eqn:ModelBalancedFourier}
\end{equation}
The two dual measurements are therefore considered to be $\pi$-phase-shifted.
In section \ref{sec:JustificationBalancedDetection}, we show that for the dual detection scheme as defined above (equation \eqref{eqn:ModelBalancedFourier}), using the equivalent matrix $\mathbf{A} = \left[
\mathbf{A_1} \hspace{0.1cm}  \mathbf{A_2}
\right]^T$ or the equivalent matrix $\mathbf{A} = 
\mathbf{A_1} - \mathbf{A_2} $ leads to approximately the same estimation variance when $M \gg 1$.
In other words, if $M \gg 1$, simply subtracting the measurements lead to a variance $\mathbf{V}_{W1b}$ approximately equal to the variance $\mathbf{V}_{W1d}$ obtained when considering the full measurement vector (equation \eqref{eqn:modelBalancedBlock}):
\begin{equation}
\mathbf{V}_{W1b} (\hat{\mathbf{x}}) \approx \mathbf{V}_{W1d} (\hat{\mathbf{x}})
\end{equation}
Below, we derive the estimation variance for the balanced-detection strategy, since it easier to derive, ans since this strategy presents a substantial advantage of computational efficiency while preserving the SNR. \\

\noindent In the balanced-detection strategy, the following relations are verified: 
\begin{equation}
\left\{
    \begin{array}{ll}
  \mathbf{A} = \mathbf{A_1} - \mathbf{A_2}  = \frac{1}{\gamma} \mathbf{W}_M \\
  \mathbf{A_1} + \mathbf{A_2} = \frac{1}{\gamma} \mathbf{J}_M
  \end{array}
  \right.
\end{equation}
Using equations \eqref{eqn:Balanced_GeneralVarReal},  \eqref{eqn:VarianceGeneralFormula_ComplxRealPart} and \eqref{eqn:W.W} leads to the estimation variance:
\begin{align*}
\mathbf{V}_{W1b}(\hat{\mathbf{x}}_{Re}^s) &  = \frac{\gamma}{2M} \bar{x}^s Re \left(
( \mathbf{W}_M^{*} \odot \mathbf{W}_M + \mathbf{W}_M^{*} \odot \mathbf{W}_M^{*}) \mathbf{1}_{M} \right) \\
&= \frac{\gamma}{2M} \bar{x}^s Re \left(
( \mathbf{J}_M + \widetilde{\mathbf{W}}_M^{*}) \mathbf{1}_{M} \right) \\
& = \frac{\gamma}{2} \bar{x}^s \mathbf{1}_M + \frac{\gamma }{2} \bar{x}^s \begin{bmatrix}
\mathbf{e_1}_{,N}\\
\mathbf{e_1}_{,N}
\end{bmatrix} 
\end{align*}
where $\widetilde{\mathbf{W}}_M^* = \mathbf{W}_M^{*} \odot \mathbf{W}_M^{*}$. 
The variance of the real part of the non-symmetric estimate $\hat{\mathbf{x}}$ then reads:
\begin{equation}
\boxed{\mathbf{V}_{W1b}(\hat{\mathbf{x}}_{Re})  = \gamma
\bar{x} (\mathbf{1}_N + 
\mathbf{e_1}_{,N})}
\end{equation}
Therefore,
\begin{equation}
V_{W1b}(\hat{x}_i)
= 
\left\{
    \begin{array}{ll}
        \gamma \bar{x} & \mbox{ } \forall i \neq 0 \\
        2 \gamma
\bar{x} & \mbox{for }  i = 0
    \end{array}
\right.
\end{equation}

\subsubsection*{Model with cosines-expressions}
\label{sec:correspondance cosine model}
Since $S(\nu) \in \mathbb{R} $ and $R(\tau)$ is hermitian, the detected intensity can also be expressed as: 
\begin{equation}
I(\tau) = \frac{1}{2\gamma} (I_0 + \int_{-\infty}^{\infty} S(\nu)cos(2\pi\nu\tau) \, \mathrm{d}\nu  )
\label{eqn:CosIntensity}
\end{equation} 
and the least-square estimation of the power spectrum is thus given by: 
\begin{equation}
\hat{S}(\nu) = \int_{-\infty}^{\infty} (2\gamma I(\tau) - I_0) cos(2\pi\nu\tau) \, \mathrm{d}\tau 
\label{eqn:CosSpectrum}
\end{equation} 
In this case, the discrete model is obtained by replacing the discrete Fourier matrix by a discrete cosine matrix of size $M \times N$, for example $\mathbf{C}_{M,N}$ with elements $c_{mn} = cos(2\pi \frac{nm}{M})$. 
Then, \eqref{eqn:CosIntensity} becomes: 
\begin{equation}
\mathbf{b} \sim Poisson (\frac{1}{2\gamma}(I_0 \mathbf{1}_M + \mathbf{C}_{M,N} \mathbf{x}) )  
\end{equation}
and \eqref{eqn:CosSpectrum}:
\begin{equation}
\hat{\mathbf{x}} = \mathbf{C}_{M,N}^{+} (2\gamma \mathbf{b} - I_0 \mathbf{1}_{M})  
\end{equation}

\subsection{Proofs for Case 1 (Fig.~\ref{fig:FTIR_Interferometer}(a))}
The same kind of proof could be derived for the matrix $\mathbf{C_1}$ defined in equation~\ref{eqn:C1_DCT_def}. We do not provide a full proof. We simply note that since the number of measurements is twice smaller than for Case 2, the variance is expected to be twice larger than for Case 2 with $\gamma = 1$. This is verified numerically in Fig.~\ref{fig:Comparison_matrices_SI_2D} and Fig.~S2.

\section{Dual detection: Justification for the balanced-strategy}
\label{sec:JustificationBalancedDetection}

In this section we show that, for dual detection (section~\ref{sec: Balanced detection}), for a subset of matrices $\mathbf{A}$, which include the Hadamard matrix $\mathbf{H}$ and the Fourier matrix $\mathbf{W}$, the estimation variance associated to the balanced detection strategy ($\mathbf{V}_{Ab}$) is approximately equal to the variance obtained when taking all the measurements ($\mathbf{V}_{Ad}$). This is important because in practise, it is the balanced detection strategy which is often used. For both Hadamard and Fourier-based multiplexing, the dual detection scheme is defined as: 
\begin{equation}
\left\{
    \begin{array}{ll}
\mathbf{A_1} + \mathbf{A_2} =  \mathbf{J_N}  \\ 
\mathbf{A_1} - \mathbf{A_2} =  \mathbf{A}  \\ 
\mathbf{A_1} = \frac{1}{2} (\mathbf{J}_N + \mathbf{A}) \\
\mathbf{A_2} = \frac{1}{2}(\mathbf{J}_N - \mathbf{A}) \\
    \end{array}
\right.
\label{eqn:balanceddet_defBIS}
\end{equation}
with $\mathbf{A} = \mathbf{H}$ or  $\mathbf{A} = \mathbf{W}$. $\mathbf{A} \in \mathbb{C}^{N^2}$.


\noindent In the following, we show that, if $N \gg 1$:
\begin{equation}
\left\{
    \begin{array}{ll}
        |a_{ij}| = 1 \\
        \mathbf{A}\mathbf{A}^H = N \mathbf{I}_N \\
        \mathbf{J}_N \mathbf{A}^{H} = N \mathbf{1} \mathbf{e_1} ^T \\
    \end{array}
\right.
\Longrightarrow 
\mathbf{V}_{Ad} (\hat{\mathbf{x}}) \approx  \mathbf{V}_{Ab} (\hat{\mathbf{x}}) \approx  \bar{x} \mathbf{1}_N 
\label{eqn:ResultApproxSubstractFullmodel}
\end{equation}
Therefore, if the matrix $\mathbf{A} \in \mathbb{C}^{N \times N}$ respects the above conditions \footnote{A matrix $\mathbf{A} \in \mathbb{C}^{N \times N}$ fulfilling the two first conditions is called a \textit{complex Hadamard matrix} \cite{Tadej2006, Banica2019}} (modulus of each entry is unity, matrix rows pairwise orthogonal, sum of all rows is zero except for the first row), both dual detections strategies lead to the same estimation variance provided $ N \gg 1$. This class of matrices include the Fourier matrix $\mathbf{W}$ and Hadamard matrix $\mathbf{H}$. Therefore, for Hadamard-based or Fourier-based positive dual-multiplexing, it is relevant to employ a balanced strategy: it is relevant to simply subtract the measurements vectors instead of considering the more complicated full model. \\ \\

\noindent \underline{Proof:} \\
When the full measurements are considered, the equivalent multiplexing matrix is: 
\begin{equation}
\mathbf{C} = \left[
\mathbf{A_1} \hspace{0.1cm} \mathbf{A_2}
\right]^T
\end{equation}
and the estimation variance in the complex case (adapted from equation~\eqref{eqn:VarianceGeneralBalanced_Block}):
\begin{equation}
\mathbf{V}_{Ad}(\mathbf{\hat{x}}) = (\mathbf{C}^{+} \odot (\mathbf{C}^{+})^*) \mathbf{Cx} 
\end{equation}
($\mathbf{C}^{H} \mathbf{C}$ is supposed not singular). Using the conditions of \eqref{eqn:ResultApproxSubstractFullmodel} and equation \eqref{eqn:Sherman_Morrison} leads to:
\begin{align*}
\mathbf{C}^{+} & = \begin{bmatrix}
\mathbf{A_1}^{H} \mathbf{A_1} + \mathbf{A_2}^{H} \mathbf{A_2}
\end{bmatrix} ^{-1}  \mathbf{C}^{H} \\
& = 2 (\mathbf{A}^{H} \mathbf{A} +  N \mathbf{J}_N )^{-1}  \mathbf{C}^{H} \\
& = \frac{2}{N} (\mathbf{I}_N  - \frac{\mathbf{J}_N}{N+1}  \mathbf{C}^{H}) \\
& = \frac{2}{N} (\mathbf{I}_N  - \frac{\mathbf{J}_N}{N+1}  ) \begin{bmatrix}
\mathbf{A_1}^{H} \hspace{0.2cm} \mathbf{A_2}^{H}
\end{bmatrix} \\
& = \frac{1}{N} \begin{bmatrix}
\frac{1}{N+1}\mathbf{J}_N + \mathbf{A}^H - \frac{1}{N+1}  \mathbf{J}_N \mathbf{A}^H \hspace{0.5cm} \frac{1}{N+1}\mathbf{J}_N - \mathbf{A}^H + \frac{1}{N+1}  \mathbf{J}_N \mathbf{A}^H
\end{bmatrix} \\
& = \frac{1}{N} \begin{bmatrix}
\frac{1}{N+1}\mathbf{J}_N + \mathbf{A}^H - \frac{N}{N+1}  \mathbf{1e_1}^T \hspace{0.5cm} \frac{1}{N+1}\mathbf{J}_N - \mathbf{A}^H + \frac{N}{N+1}  \mathbf{1e_1}^T
\end{bmatrix}
\end{align*}
Replacing in \eqref{eqn:VarianceGeneralBalanced_Block} leads to:
\begin{align*}
\mathbf{V}_{Ad}  & = \scriptstyle \frac{1}{N^2(N+1)} \begin{bmatrix} \frac{1+(N+1)^2}{(N+1)}\mathbf{J}_N + \mathbf{A}^H + \mathbf{A}^T - \frac{N(N+4)}{N+1}\mathbf{1e_1}^T
\hspace{0.5cm} 
\frac{1+(N+1)^2}{(N+1)}\mathbf{J}_N -\mathbf{A}^H + \mathbf{A}^T - \frac{N^2}{N+1}\mathbf{1e_1}^T
\end{bmatrix}
\begin{bmatrix}
\mathbf{A_1} \\
\mathbf{A_2} 
\end{bmatrix}
\mathbf{x} \\
&= \frac{1}{N^2} \left( N (1 - \frac{N+3}{(N+1)^2}) \mathbf{J}_N + \frac{1}{N+1}(\mathbf{A}^H + \mathbf{A}^T) \mathbf{A} \right) \mathbf{x} \\
&=  \bar{x} (1 - \frac{N+3}{(N+1)^2}) \mathbf{1}_N + \frac{1}{N(N+1)}(1 + \frac{1}{N}\mathbf{A}^T\mathbf{A}) \mathbf{x}  
\end{align*}
Therefore, if $N \gg 1$, and since $|a_{ij}| = 1$, the variance obtained with the complete measurement reads: 
\begin{equation}
\mathbf{V}_{Ad}(\hat{\mathbf{x}})  \approx  \bar{x} \mathbf{1}_N 
\end{equation} \\

On the other hand, when subtracting the measurements, the equivalent multiplexing matrix is 
\begin{equation}
\mathbf{A_1}-\mathbf{A_2} = \mathbf{A}
\end{equation}
and the associated estimation variance is (adapted from equation~\eqref{eqn:Balanced_GeneralVarReal}) is: 
\begin{equation}
\mathbf{V}_{Ab}(\hat{\mathbf{x}}) = N\bar{x}(\mathbf{A}^{-1} \odot (\mathbf{A}^{-1})^*) \mathbf{1}_N
\end{equation}
Then, using the properties on the left of equation~\eqref{eqn:ResultApproxSubstractFullmodel}, the variance reads:
\begin{equation}
\mathbf{V}_{Ab}(\hat{\mathbf{x}}) = N \bar{x}\frac{1}{N^2}\mathbf{J}_N \mathbf{1}_N =\bar{x}\mathbf{1}_N 
\end{equation}



\section{Numerical results}
Here, we support and illustrate the theoretical proofs with some simulations in 1-D and 2-D.  

\begin{figure} [H]
\centering
\textbf{\includegraphics[scale=0.7]{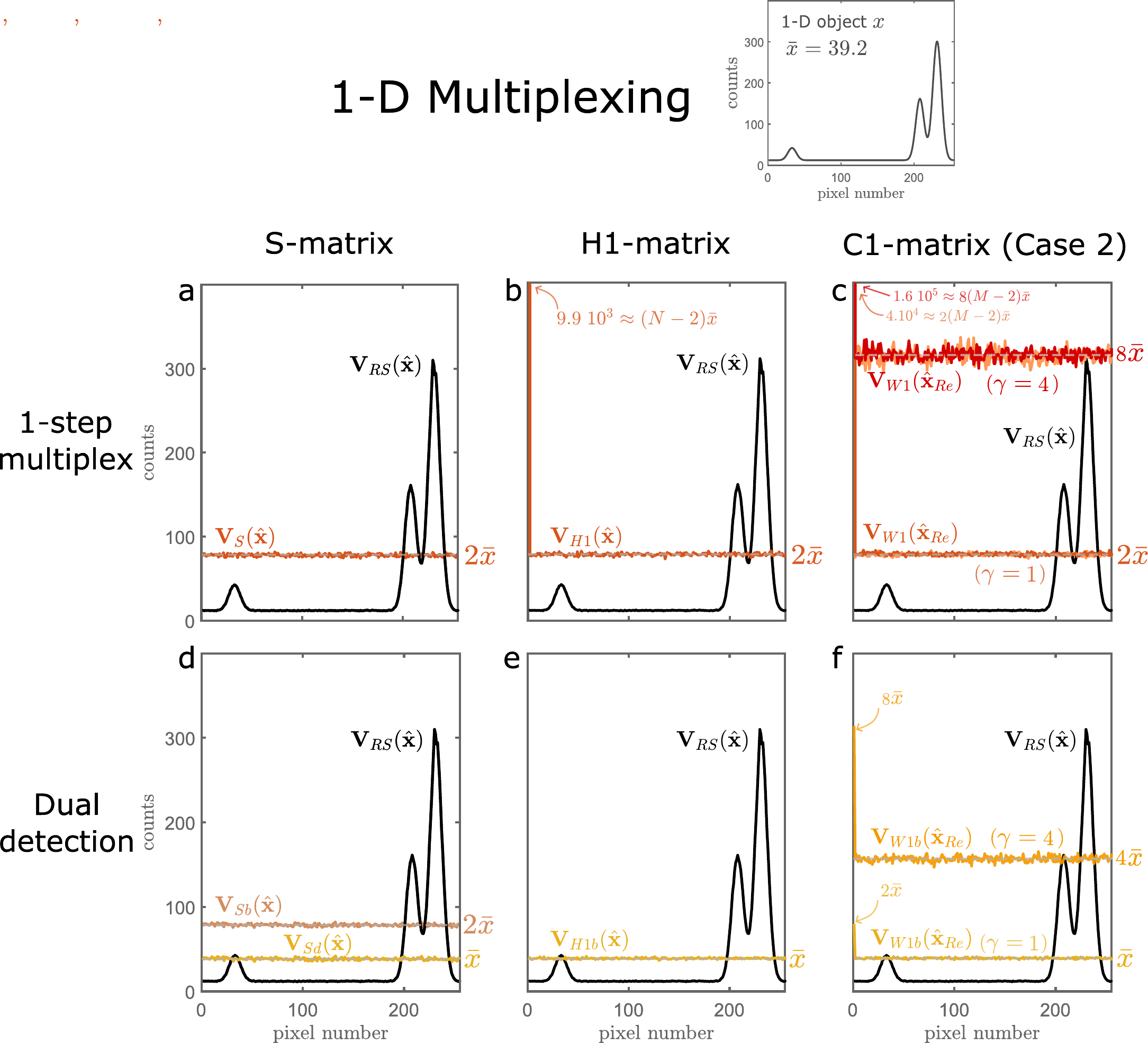} }
\caption{Empirical variances obtained in simulations with LS-estimation, for 1-step multiplexing and dual-detection, for the 3 considered multiplexing matrices, as compared to the variance associated with raster-scanning (black curves). For positive-cosine multiplexing, we show the resulting variance for $\gamma=1$ and $\gamma=4$. (c) For positive-cosine multiplexing, the variance superimpose with the results obtained when multiplexing with a positive cosine matrix (light orange curves, see section~\ref{sec:correspondance cosine model}). (d): For S-multiplexing, we find that the variance obtained with a balanced strategy ($V_{Sb}(\mathbf{\hat{x}})$) is twice higher than when considering the full measurements ($V_{Sd}(\mathbf{\hat{x}})$). $N = 256$, $M = 510$, number of noise realisations $ = 5000$.}
\label{fig:Comparison_matrices_SI_1D}
\end{figure}

\begin{figure} [H]
\centering
\textbf{\includegraphics[scale=0.7]{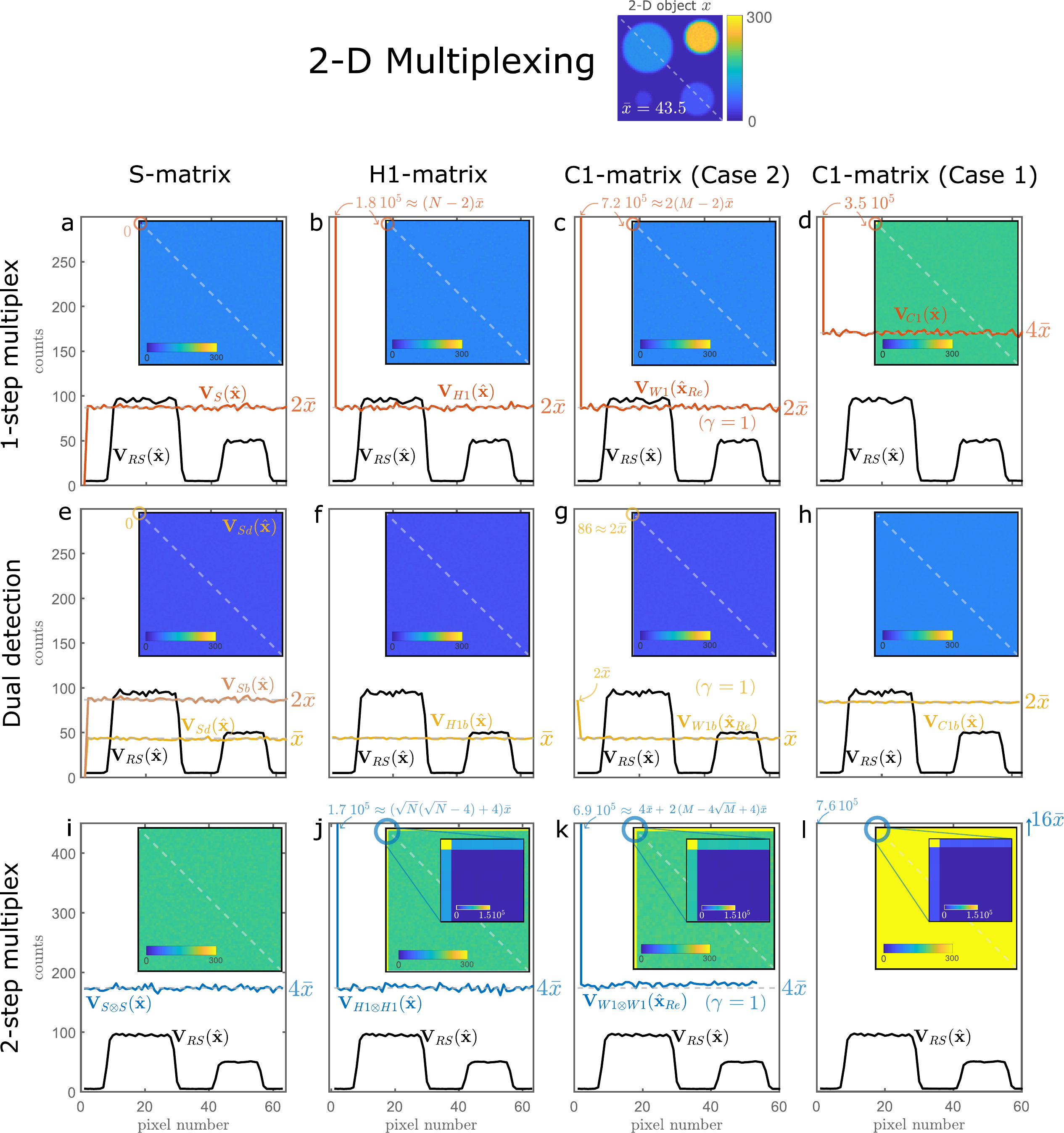} }
\caption{Empirical variances obtained in simulations with LS-estimation, for 1-step multiplexing, dual-detection and 2-step multiplexing, for several multiplexing matrices, as compared to the variance associated with raster-scanning (black curves). The images represent the 2-D variance along the pixels of the 2-D intensity object. The plots are the sections along the image diagonal. In (a) and (e), the first pixel variance is equal to zero, because we need to constrain the first object pixel to be zero to use the S-matrix of odd dimensions in 2-D (see paragraph below). In (e), the variance obtained with balanced detection is twice higher than when considering the full measurements. In (j-l), we zoom of the special pixels of the variance introduced by the structure of $\mathbf{H_1}$ and $\mathbf{C_1}$ . $N = 4096$, $M = 8190$, number of noise realisations $ = 5000$.} 
\label{fig:Comparison_matrices_SI_2D}
\end{figure}

\noindent Some remarks on the implementation in 1-D and 2-D:
\begin{itemize}
\item The two-step multiplexing modality is not relevant in 1-D
\item In 2-D, for one-step multiplexing, the 2-D patterns are obtained by reshaping each of rows of the multiplexing matrix. But in this work, the S-matrices are of odd dimensions (derived from Sylvester-Hadamard matrices). The S-matrix can be used in 2-D with a 'negative' Hadamard matrix $\mathbf{H_2}$ \eqref{eqn:def_H2}), if the first pixel of the object is zero and if the first measurement is discarded. 
\end{itemize}


\chapter{Robustness to perturbations}
\label{SI_otherBckmodels}

In this chapter, we assess the robustness of the system to some perturbations that modify the initial equation~\eqref{eqn:completemodel}:
\begin{align*}
\mathbf{b} \sim Poisson(\mathbf{Ax}) 
\end{align*} 

\section{Initial model multiplied by a constant}
\label{sec:Perturb_Gamma}
Here, we consider a perturbation $\epsilon > 0 $ that modify the initial system such that: 
\begin{equation}
\mathbf{b} \sim Poisson(\frac{1}{\epsilon}\mathbf{A} \mathbf{x} )
\label{eqn:PertubModelEpsilon}
\end{equation}
This accounts for various system losses or specific experimental details.  \\

\noindent Often, this perturbation is not quantified and is simply incorporated in the object. Then, the quantity of interest is no longer the ground truth $\mathbf{x}$ but a quantity proportional to it: $\mathbf{y}=\frac{1}{\epsilon} \mathbf{x}$. 
With this change of variable of interest, the initial model applies with
$\mathbf{b} \sim Poisson(\mathbf{A} \mathbf{y} )$, $\mathbf{\hat{y}} = \mathbf{A}^{-1} \mathbf{b}$, and $\mathbf{V}(\hat{\mathbf{y}}) = 
(\mathbf{A}^{-1}\odot\mathbf{A}^{-1})\mathbf{A}\mathbf{y} = \epsilon \mathbf{V}(\hat{\mathbf{x}})$. \\

\noindent The perturbation can also be seen to impact the initial multiplexing matrix such that it is no longer $\mathbf{A}$ but rather an equivalent matrix $\widetilde{\mathbf{A}}$. In this case, equation~\eqref{eqn:PertubModelEpsilon} becomes:
\begin{equation}
\mathbf{b} \sim Poisson(\widetilde{\mathbf{A}} \mathbf{x} )
\end{equation}
with 
\begin{equation}
\widetilde{\mathbf{A}} = \frac{1}{\epsilon} \mathbf{A}
\end{equation}
Using \eqref{eqn:VarianceGeneralFormula_Real} directly leads to the associated estimation variance: 
\begin{equation}
\widetilde{\mathbf{V}}(\hat{\mathbf{x}})
 = (\widetilde{\mathbf{A}}^{-1}\odot\widetilde{\mathbf{A}}^{-1})\widetilde{\mathbf{A}}\mathbf{x} = \epsilon (\mathbf{A}^{-1}\odot\mathbf{A}^{-1})\mathbf{A}\mathbf{x}
\end{equation}
i.e.
\begin{equation}
\boxed{\widetilde{\mathbf{V}}(\hat{\mathbf{x}}) = \epsilon \mathbf{V}(\hat{\mathbf{x}})}
\label{eqn:pertubEpsilon_ResultGlobal}
\end{equation}
Where $\mathbf{V}(\hat{\mathbf{x}})$ is the variance associated to $\mathbf{A}$ given in equation \eqref{eqn:VarianceGeneralFormula_Real}.

\subsection{Implications for the three multiplexing schemes}
The implications for of this result for the three positive-multiplexing schemes are listed in the table below. 
{\renewcommand{\arraystretch}{2}
\begin{table} [H]
\centering
\begin{tabularx}{\linewidth}{|p{2cm}|X|X|X|X|}
\hline
 &  One-step multiplexing & \multicolumn{2}{|c|}{Two-step multiplexing}  & Dual-detection\\
\hline
$\widetilde{\mathbf{A}}$ &  $\frac{1}{\epsilon}\mathbf{A}$ & $\frac{1}{\epsilon}\mathbf{A}$ & $\frac{1}{\epsilon_p}\mathbf{P}^T \otimes \frac{1}{\epsilon_u} \mathbf{U}$  & $\frac{1}{\epsilon} \begin{bmatrix}
\mathbf{A_1} \hspace{0.1cm} \mathbf{A_2}
\end{bmatrix}^T$ \\ 
\hline
$\widetilde{\mathbf{V}}(\hat{\mathbf{x}})$ &  $\epsilon \mathbf{V}(\hat{\mathbf{x}})$  & $\epsilon \mathbf{V}(\hat{\mathbf{x}})$ & $\epsilon_p \epsilon_u \mathbf{V}(\hat{\mathbf{x}})$   & $\epsilon \mathbf{V}(\hat{\mathbf{x}})$   \\
\hline
\end{tabularx}
\caption{Variances for different perturbations of the multiplexing matrix.}
\label{tab:PertubEpsilon}
\end{table}

\subsubsection*{Implications for one-step multiplexing}
In a one-step multiplexing scenario, the above equation~\eqref{eqn:pertubEpsilon_ResultGlobal} directly applies. 

\subsubsection*{Implications for two-step multiplexing}
\label{sec:Perturb_Gamma_2steps}
In a two-step multiplexing scheme (section~\ref{sec: Two-step multiplexing}), the perturbation may happen on the global multiplexing matrix $\mathbf{A} = \mathbf{P}^T \otimes \mathbf{U}$, in which case the above result \eqref{eqn:pertubEpsilon_ResultGlobal} applies.
On opposite, if two perturbations are applied on each sub-matrix such that: $\widetilde{\mathbf{P}} = \frac{1}{\epsilon_p} \mathbf{P}$ and  $\widetilde{\mathbf{U}} = \frac{1}{\epsilon_u} \mathbf{U}$, the equivalent multiplexing matrix is: 
\begin{equation}
\widetilde{\mathbf{A}} = \frac{1}{\epsilon_p \epsilon_u } \mathbf{A}
\end{equation}
Hence the resulting variance reads: 
\begin{equation}
\widetilde{\mathbf{V}}(\hat{\mathbf{x}})
 = \epsilon_p \epsilon_u  \mathbf{V}(\hat{\mathbf{x}})
\label{eqn:pertubGamma_Result2steps}
\end{equation}
Where $\mathbf{V}(\hat{\mathbf{x}})$ is the variance associated to $\mathbf{A}$ given in equation \eqref{eqn:Kron_GeneralFormula}.

\subsubsection*{Implications for dual-detection}
\label{sec:Perturb_Gamma_Dual}
In the dual detection modality, the perturbation can modelled via the following equivalent matrix: 
\begin{equation}
\widetilde{\mathbf{A}} =  \frac{1}{\epsilon} \begin{bmatrix}
\mathbf{A_1} \hspace{0.1cm}  \mathbf{A_2}
\end{bmatrix} ^T = \frac{1}{\epsilon} \mathbf{A}
\end{equation}
In this case we also have: 
\begin{equation}
\widetilde{\mathbf{V}}(\hat{\mathbf{x}})
 = \epsilon \mathbf{V}(\hat{\mathbf{x}})
 \label{eqn:pertubGamma_ResultDual}
\end{equation}
Where $\mathbf{V}(\hat{\mathbf{x}})$ is the variance associated to $\mathbf{A}$ given in equation \eqref{eqn:VarianceGeneralBalanced_Block}.\\

\noindent \underline{Proof}:
Adapting equation \eqref{eqn:VarianceGeneralBalanced_Block} yields: 
\begin{equation}
\mathbf{V}(\hat{\mathbf{x}}) = (\widetilde{\mathbf{A}}^{+} \odot \widetilde{\mathbf{A}}^{+}) \widetilde{\mathbf{A}} \mathbf{x}
\end{equation}
with
\begin{equation}
\widetilde{\mathbf{A}}^{+} = (\widetilde{\mathbf{A}}^{T}\widetilde{\mathbf{A}}^{T})^{-1}\widetilde{\mathbf{A}}^{T} = \left(\frac{1}{\epsilon^2} \mathbf{A}^T \mathbf{A} \right)^{-1} \times \frac{1}{\epsilon} \mathbf{A}^T
= \gamma \left( \mathbf{A}^T \mathbf{A} \right) \mathbf{A}^T = \epsilon \mathbf{A}^+
\end{equation}
Combining the two above equations leads to the results of equation \eqref{eqn:pertubGamma_ResultDual}.


\subsection{Implications for comparisons at constant number of photons}
\label{sec:SI_constantNbPhotons}

In this work, we focus on comparing positive-multiplexing and raster-scanning when the number of photons between is \textit{not} constant. Then, for $N$ measurements, raster-scanning leads to a total of $N \bar{x}$ photon counts, and positive-multiplexing to a total of about $\frac{N^2}{a} \bar{x}$ photon counts (where $a$ depends on the multiplexing matrix). \\
Yet, the above result of equation \eqref{eqn:pertubEpsilon_ResultGlobal} also allows us to compare the variance of raster-scanning and positive-multiplexing at constant number of photons (same number of photons detected in the two cases). 
Comparing raster-scanning and positive-multiplexing at constant number of photons is equivalent to divide the number of collected photons by multiplexing $\frac{N}{a}$ and thus multiplex with an equivalent matrix $\widetilde{\mathbf{A}} = \frac{a}{N} \mathbf{A}$. Then, adapting equation~\ref {eqn:pertubEpsilon_ResultGlobal} results in a positive-multiplexing variance of:
\begin{equation}
\widetilde{\mathbf{V}}_{mult}(\hat{\mathbf{x}}) = \frac{N}{a} \mathbf{V}(\hat{\mathbf{x}})
\end{equation}
where $\mathbf{V}(\hat{\mathbf{x}})$ is the variance when the number of photons is $\frac{N^2}{a} \bar{x}$.
The variance associated with raster-scanning stays $\mathbf{V}_{rs}(\hat{\mathbf{x}}) = \mathbf{x}$. 
\\

\noindent Therefore, performing positive-multiplexing with fewer photons worsen the associated variance by the same factor: the estimation variance is worsened by a factor proportional to $N$, and thus the SNR by a factor proportional $\sqrt{N}$. \\

\noindent Note that in positive-Hadamard-based and positive-Cosine multiplexing, $a=2$ and  $\mathbf{V}(\hat{\mathbf{x}})= k\bar{x}$, therefore $\widetilde{\mathbf{V}}_{mult}(\hat{\mathbf{x}}) = \frac{N}{2}k\bar{x}$. 



\section{Initial model with additional noise or background}
In this section, we consider some additional noise sources or unwanted background signal that may arise in the experimental system (Fig.~\ref{fig:bloc_diagram}):
\begin{itemize}
\item Additional white Gaussian noise (AWGN)  $\mathbf{e}$ at the detection (e.g. electronic noise arising from the detector)
\item Additional known background $\bm{\eta}$ that does not experience the multiplexing step
\item Additional known background $\bm{\beta}$  that does experience the multiplexing step
\item Additional constant offset $\alpha$ in the multiplexing matrix
\end{itemize}



\begin{figure} [H]
\centering
\textbf{\includegraphics[scale=0.7]{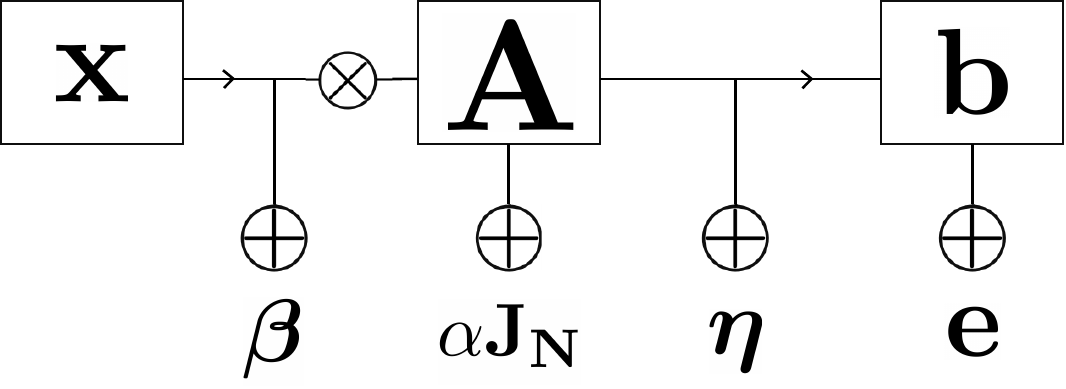} }
\caption{Schematic representation of some sources of noise or unwanted background that may arise during a measurement. $\mathbf{x}$: object, $\mathbf{A}$: multiplexing matrix, $\mathbf{b}$: measurement, $\bm{\beta}$: unwanted background or signal added to the system before the multiplexing stage, $\alpha \mathbf{J}_N$: unwanted constant offset added to the multiplexing matrix, $\bm{\eta}$: unwanted background or signal added to the system after the multiplexing stage, $\mathbf{e}$: detector electronic noise}
\label{fig:bloc_diagram}
\end{figure}

\noindent For each scenario, we derive the variances associated to raster-scanning, and to positive-multiplexing with a general matrix $\mathbf{A}$. 
We also give the results for a special class of matrices $\mathbf{A_c}$ that lead to a constant variance: 
\begin{equation*}
\mathbf{V}_{Ac}(\mathbf{\hat{x}}) = k \bar{x} \mathbf{1}_N 
\end{equation*} 
Details on this class of matrices are provided in chapter \ref{sec: SI_Generalisation}. $\mathbf{A_c}$ is defined such that  $\mathbf{A_c}\in \mathbb{R}^{N^2}$ (for $N >2$) with an inverse of: 
\begin{align*}
\mathbf{A_c}^{-1} = \frac{k}{N} (N-2p) \mathbf{\Lambda}
\end{align*}
where $\mathbf{\Lambda}$ is a matrix with elements $\delta_{ij} = \pm 1$, in which each column contains $p$ negative elements (equation~\eqref{eqn:SufficientConditionAinv}). 
\subsection{Results}
The obtained variance for each scenario are summarized in the table below. The proofs are provided in section \ref{sec: SI_OtherBckModel_ProofsDetails}. The results of the table are illustrated in Fig.~\ref{fig:BckModels} for S-multiplexing ($\mathbf{A_c} = \mathbf{S}-$matrix, i.e.  $k=\frac{2N}{N+1}\approx 2$ and $q=1$). 

{\renewcommand{\arraystretch}{2}
\begin{table} [H]
\centering
\begin{tabularx}{\linewidth}{|p{1.3cm}|p{2.9cm}|X|X|p{3.4cm}|}
\hline
& AWGN $\mathbf{e}$ & Known bck $\bm{\eta}$ & Known bck $\bm{\beta}$ & Matrix with constant offset $\alpha$ \\ 
\hline
$\mathbf{b_0}$ &  \scalebox{0.9}{$\mathbf{Ax}$} & \scalebox{0.9}{$\mathbf{Ax} + \bm{\eta} $} & \scalebox{0.9}{$\mathbf{A} (\mathbf{x} + \bm{\beta})$} & \scalebox{0.9}{$(\mathbf{A} + \alpha \mathbf{J_N} ) \mathbf{x}$}  \\
$\mathbf{b}$ &  \scalebox{0.9}{$Poisson(\mathbf{Ax}) + \mathbf{e}$} & \scalebox{0.9}{$Poisson(\mathbf{Ax} + \bm{\eta}) $} & \scalebox{0.9}{$Poisson(\mathbf{A} (\mathbf{x} + \bm{\beta}))$} & \scalebox{0.9}{$Poisson((\mathbf{A} + \alpha \mathbf{J}_N ) \mathbf{x}$)} \\
\hline
$\hat{\mathbf{x}}_{LS}$ &  \scalebox{0.9}{$\mathbf{A}^{-1} \mathbf{b}$} & \scalebox{0.9}{$\mathbf{A}^{-1} (\mathbf{b} - \bm{\eta} )$} & \scalebox{0.9}{$\mathbf{A}^{-1} \mathbf{b} - \bm{\beta} $} & \scalebox{0.9}{$(\mathbf{A} + \alpha \mathbf{J_N} )^{-1} \mathbf{b}$} \\ 
\hline
\multicolumn{5}{|c|}{General case} \\
\hline
\scalebox{0.9}{$\widetilde{\mathbf{V}}(\hat{\mathbf{x}})$} & \scalebox{0.9}{$\mathbf{V}(\hat{\mathbf{x}}) +$} \scalebox{0.9}{$\sigma^2 diag((\mathbf{A}^T \mathbf{A})^{-1})$} & \scalebox{0.9}{$\mathbf{V}(\hat{\mathbf{x}}) + (\mathbf{A}^{-1} \odot \mathbf{A}^{-1})\bm{\eta}$} & \scalebox{0.9}{$ \mathbf{V}(\hat{\mathbf{x}}) + (\mathbf{A}^{-1} \odot \mathbf{A}^{-1})\mathbf{A}\bm{\beta}$} & \scalebox{0.9}{$ ((\mathbf{A} + \alpha \mathbf{J_N} )^{-1})\odot  (\mathbf{A} + $} \scalebox{0.9}{$\alpha \mathbf{J_N} )^{-1}) (\mathbf{A} + \alpha \mathbf{J_N} ) \mathbf{x}$} \\ 
\hline
\multicolumn{5}{|c|}{Raster-scanning} \\
\hline
\scalebox{0.9}{$\widetilde{V}_{RS}(\hat{x}_i) $} &  $x_i  +  \sigma^2 $ & $x_i + \eta_i $  & $x_i + \beta_i $ & $x_i + \alpha N \bar{x}$ \scalebox{0.8}{(if $\alpha \ll 1$)}\\ 
\hline
\multicolumn{5}{|c|}{Positive-multiplexing with $\mathbf{A} = \mathbf{A_c}$ (such that $V_{Ac} (\hat{x}_i)= k\bar{x} $)} \\
\hline
\scalebox{0.9}{$\widetilde{V}_{Ac} (\hat{x}_i)  $} &  $k \bar{x}  +  \frac{k^2 q^2}{N}\sigma^2$ & $k \bar{x} + \frac{k^2 q^2}{N} \bar{\eta} $  & $k \bar{x} + k \bar{\beta}$ & $k \bar{x} + \alpha k^2  q^2 \bar{x} $ \scalebox{0.8}{(if $N \gg 1$)} \\
\hline
\end{tabularx}
\caption{Resulting variances in the presence of noise or unwanted signal.  AWGN: Additive White Gaussian Noise; $\sigma$: standard deviation of $\mathbf{e}$; $\mathbf{b_0}$: general noiseless model, $\mathbf{b}$: noise model, $\hat{\mathbf{x}}_{LS}$: least-square estimate,$\mathbf{V}(\hat{\mathbf{x}})$: associated variance. $k$: real positive constant. $q = (N - 2p)$: defined in equation~\eqref{eqn:SufficientConditionAinv}.
$\eta_i \geq 0$; $\beta_i \geq 0$ and $\alpha$ are supposed to be known from a calibration step.
The proofs and exact expressions are provided in the text.}
\label{tab:Four scenarios}
\end{table}

\noindent Overall, raster-scanned measurements are highly sensitive to the four considered additional perturbations: the additional noise variance adds to the signal.
In opposite, in the two first scenarios, positive-multiplexing with $\mathbf{Ac}$ is very robust to the perturbations $\mathbf{e}$ and $\bm{\eta}$: they hardly impact the variance if $N \gg 1$. The stronger the perturbation, the more object pixels are better estimated with Ac-multiplexing (as compared to raster-scanning). Note that in the limit case where $\sigma^2 \geq k \bar{x}$ or $\eta_i \geq k \bar{x}$, Ac-multiplexing brings a SNR advantage over raster-scanning on all pixels. The effect is similar in the last scenario. 
The results are different in the third scenario, i.e. if some background signal $\bm{\beta}$ is added to the system before the multiplexing step. 
There, the perturbation impacts on average $k$ times more Ac-multiplexing than raster-scanning. Then, the stronger the perturbation,
the more object pixels are better estimated with  raster-scanning (as compared to Ac-multiplexing).

\begin{figure} [H]
\centering
\textbf{\includegraphics[width=\linewidth]{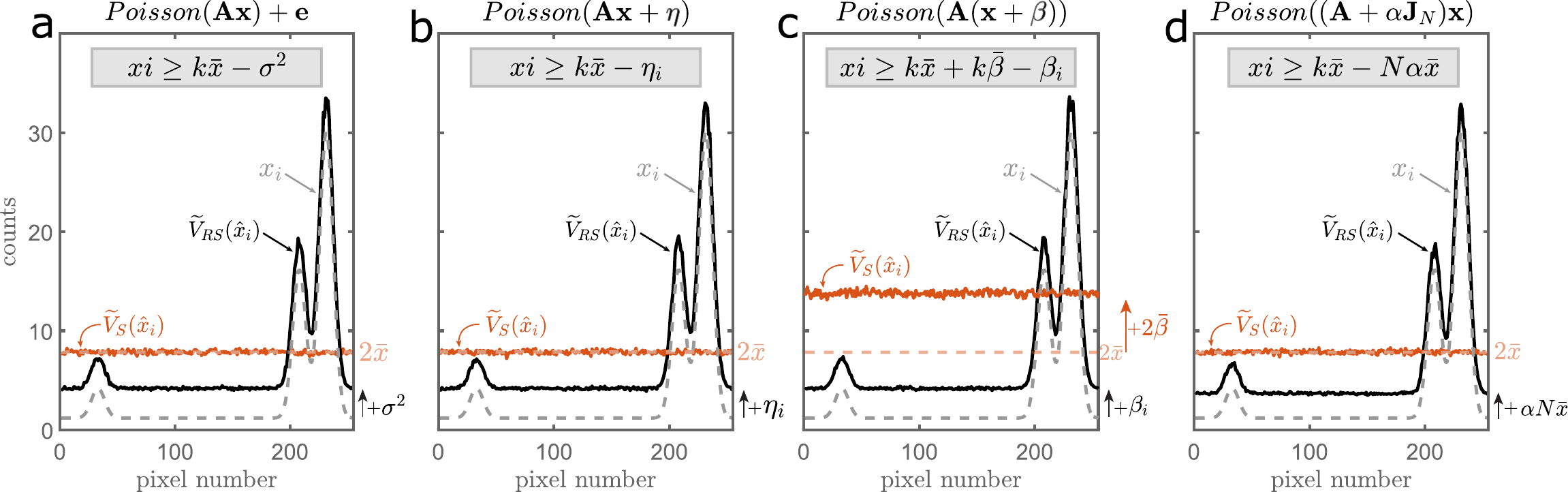} }
\caption{Empirical variances obtained for raster-scanning (black) and positive multiplexing with $\mathbf{A_c} = \mathbf{S}$ (\textcolor{red}), for the four above scenarios and a 1-D object. For comparison, the theoretical variances that would be obtained in the initial shot-noise limited model (equation~\eqref{eqn:completemodel}) are indicated in grey and light red dashed lines for raster-scanning and S-multiplexing respectively. The gray boxes indicate the conditions on object pixel $i$ for Ac-multiplexing to bring an SNR advantage over raster-scanning (valid for for $N \gg 1$, $\alpha \ll 1$ and $q$ independent of $N$). In the simulation, $\eta$ and $\beta$ are constant backgrounds, $\sigma^2 = \eta_i = \beta_i = 3$ counts; $N\alpha \bar{x}=2.6$ counts; with $\alpha=0.0025$, $\bar{x} = 4$ counts; $N = 255$ pixels. Number of noise realisations $=$ 5000.}
\label{fig:BckModels}
\end{figure}

\subsection{Details and proofs} 
\label{sec: SI_OtherBckModel_ProofsDetails}

\subsubsection*{Additional electronic noise $\bm{e}$}
\label{subsec:Mixture of photon noise and electronic noise}
With additive electronic noise, the model becomes:
\begin{equation} 
\mathbf{b} \sim Poisson(\mathbf{Ax}) + \mathbf{e}
\end{equation}
The electronic noise is modelled as additive white gaussian noise (AWGN) $\mathbf{e}$, with $\forall i \neq j$, $E(e_i) = 0$, $E(e_ie_j) = 0$, $E(e_i^2) = \sigma^2$, where $\sigma$ is the noise standard deviation. \\
To simplify the calculation, we re-write $\mathbf{b}$ as
\begin{equation} 
\mathbf{b} = \mathbf{p} + \mathbf{e}
\end{equation}
where $\mathbf{p}$ approximates the Poisson distribution.  $\mathbf{p}$ follows a Gaussian distribution with mean equals its variance $E(p_i) = E(p_i^2) = (\mathbf{Ax})_i$. 
The error $\delta\hat{\mathbf{x}}$ reads:
\begin{equation}
\delta\hat{\mathbf{x}} = \mathbf{A}^{-1}(\mathbf{p}+\mathbf{e})
\end{equation}
This leads to the covariance matrix:
\begin{align*}
\mathbf{\Gamma} & = \mathbf{A}^{-1} (\langle\mathbf{p}\mathbf{p}^T\rangle + \langle \mathbf{e}\mathbf{e}^T\rangle) \mathbf{A}^{-T} \text{ \small (uncorrelated noise)} \\
& = \mathbf{A}^{-1} (Diag(\mathbf{Ax}) + \sigma^2 \mathbf{I}_N)) \mathbf{A}^{-T} 
\end{align*}
The resulting variance is then:
\begin{equation}
\boxed{\widetilde{\mathbf{V}}(\hat{\mathbf{x}}) = (\mathbf{A}^{-1} \odot \mathbf{A}^{-1}) \mathbf{A} \mathbf{x} + \sigma^2 diag( (\mathbf{A}^{T} \mathbf{A})^{-1})}
\end{equation}
Hence, for raster-scanning, the variance reads:
\begin{equation}
\boxed{\widetilde{\mathbf{V}}_{RS} (\hat{\mathbf{x}})= \mathbf{x}  +  \sigma^2   \mathbf{1}_N}
\end{equation}
and for multiplexing with $\mathbf{A_c}$, the variance reads: 
\begin{align*}
\widetilde{\mathbf{V}}_{Ac}(\hat{\mathbf{x}})  = k \bar{x} \mathbf{1}_N + \sigma^2 diag( (\mathbf{A_c}^{T} \mathbf{A_c})^{-1}) 
\end{align*}
\begin{equation}
\boxed{\widetilde{\mathbf{V}}_{Ac}(\hat{\mathbf{x}})  = ( k \bar{x} +  \frac{k^2 q^2}{N} \sigma^2 ) \mathbf{1}_N}
\end{equation}
with $q = (N - 2p)$. 
If $q$ is independent of $N$ (true for matrices $\mathbf{S}$, $\mathbf{H_1}$ and $\mathbf{C_1}$) and $N \gg 1$: 
\begin{align*}
\widetilde{\mathbf{V}}_{Ac}(\hat{\mathbf{x}}) \approx k \bar{x} \mathbf{1}_N
\end{align*}
Then, under these assumptions positive-multiplexing with $\mathbf{A_c}$ is advantageous over raster-scanning on pixels $i$ for which: 
\begin{align*}
\boxed{ x_i \geq  k \bar{x}  -  \sigma^2 }
\end{align*}
\underline{Example for the $\mathbf{S}-$matrix}:  $k = 2N/(N+1)$ and $q =1$, thus:
\begin{align*}
\mathbf{V}_{S} (\hat{\mathbf{x}})
& = \frac{2N}{N+1} \bar{\mathbf{x}} + \frac{4N}{(N+1)^2}  \mathbf{1}_N 
\approx ( 2 \bar{x}  +  \frac{4\sigma^2}{N}  ) \mathbf{1}_N \approx  2 \bar{x} \mathbf{1}_N \text{ (for } N \gg 1)
\end{align*}


\subsubsection*{Additional non-multiplexed known background $\bm{\eta}$}
\label{non-multiplexed known background}
This case resembles the above case. The background does not experiences the multiplexing matrix. It can depend on the object or not, and is supposed to be known from a calibration step. The model is:
\begin{equation} 
\mathbf{b} \sim Poisson(\mathbf{A} \mathbf{x} + \bm{\eta} )
\end{equation}
The error $\delta\hat{\mathbf{x}}$ for the LS estimate reads:
\begin{equation}
\delta\hat{\mathbf{x}} = \mathbf{A}^{-1}(\mathbf{b}-\bm{\eta})  - \mathbf{A}^{-1}(\mathbf{b_0}-\bm{\eta}) = \mathbf{A}^{-1} \delta\mathbf{b}
\end{equation}
and leads to the covariance matrix: 
\begin{equation}
\mathbf{\Gamma} =  \mathbf{A}^{-1} Diag(\mathbf{A} \mathbf{x} + \bm{\eta} ) \mathbf{A}^{-T}
\end{equation}
Thus:
\begin{equation}
\boxed{\widetilde{\mathbf{V}}(\hat{\mathbf{x}}) = (\mathbf{A}^{-1} \odot \mathbf{A}^{-1})\mathbf{A}\mathbf{x} + (\mathbf{A}^{-1} \odot \mathbf{A}^{-1})\bm{\eta)} }
\end{equation}
For raster-scanning, 
\begin{equation}
\boxed{\widetilde{\mathbf{V}}_{RS} (\hat{\mathbf{x}}) = \mathbf{x} + \bm{\eta} }
\end{equation}
For multiplexing with $\mathbf{A_c}$, the variance reads: 
\begin{align*}
\widetilde{\mathbf{V}}_{Ac}(\hat{\mathbf{x}}) & = k \bar{x} \mathbf{1}_N +  \frac{k^2}{N^2} (N-2 p)^2 \mathbf{J}_N \bm{\eta} 
\end{align*}
\begin{equation}
\boxed{\widetilde{\mathbf{V}}_{Ac}(\hat{\mathbf{x}}) = ( k \bar{x}  +  \frac{k^2 q^2}{N}  \bar{\eta}  ) \mathbf{1}_N}
\end{equation}
with $q = (N-2p)$. If $q^2$ is independent of $N$ (true for matrices $\mathbf{S}$, $\mathbf{H_1}$ and $\mathbf{C_1}$) and $N \gg 1$: 
\begin{equation}
\widetilde{\mathbf{V}}_{Ac}(\hat{\mathbf{x}}) \approx k \bar{x} \mathbf{1}_N
\end{equation}
In this case, multiplexing with $\mathbf{A_c}$ is advantageous over raster-scanning for pixels $i$ for which: 
\begin{equation}
\boxed{x_i \geq  k \bar{x}  - \eta_i  }
\end{equation}
\underline{Example for the $\mathbf{S}-$matrix}: 
$k = 2N/(N+1)$ and $q =1$, thus:
\begin{align*}
\widetilde{\mathbf{V}}_S (\hat{\mathbf{x}})
= \frac{2N}{N+1} \bar{x} \mathbf{1}_N + \frac{4N}{(N+1)^2} \bar{\eta} \mathbf{1}_N 
\approx (2 \bar{x} + \frac{4}{N} \bar{\eta} ) \mathbf{1}_N \approx 2 \bar{x} \mathbf{1}_N  \hspace{0.2cm} \text{ (for } N \gg 1)
\end{align*}


\subsubsection*{Additional multiplexed known background $\bm{\beta}$}
\label{multiplexed background}
Here, there is again some known background that may arise from different experimental sources. It experiences the multiplexing matrix, which means it is added to the system before the multiplexing step. It may again depend on the object or not. The model reads:
\begin{equation} 
\mathbf{b} \sim Poisson(\mathbf{A} (\mathbf{x} + \bm{\beta} ))
\end{equation}
The error $\delta\hat{\mathbf{x}}$ reads:
\begin{equation}
\delta\hat{\mathbf{x}} = \mathbf{A}^{-1}\mathbf{b}-\bm{\beta}  - (\mathbf{A}^{-1}\mathbf{b_0}-\bm{\beta}) = \mathbf{A}^{-1} \delta\mathbf{b}
\end{equation}
This leads to the covariance matrix:
\begin{equation}
\mathbf{\Gamma}  = \mathbf{A}^{-1} Diag(\mathbf{A} (\mathbf{x} + \bm{\beta} )) \mathbf{A}^{-T}
\end{equation}
and to the variance:
\begin{equation}
\boxed{\widetilde{\mathbf{V}}(\hat{\mathbf{x}}) = (\mathbf{A}^{-1} \odot \mathbf{A}^{-1})\mathbf{A}\mathbf{x} +(\mathbf{A}^{-1} \odot \mathbf{A}^{-1})\mathbf{A} \bm{\beta} }
\end{equation}
For raster-scanning: 
\begin{equation}
\boxed{\widetilde{\mathbf{V}}_{RS}(\hat{\mathbf{x}}) = \mathbf{x} + \bm{\beta} }
\end{equation}
For multiplexing with $\mathbf{A_c}$, the variance reads: 
\begin{equation}
\boxed{\widetilde{\mathbf{V}}_{Ac} (\hat{\mathbf{x}}) = (k \bar{x} + k \bar{\beta}) \mathbf{1}_N}
\end{equation}
Therefore, here multiplexing with $\mathbf{A_c}$ is advantageous over raster-scanning for pixels $i$ for which: 
\begin{equation}
\boxed{x_i \geq  k \bar{x}  +  k\bar{\beta} - \beta_i   }
\end{equation}
Or, if $\beta_i = \bar{\beta} \forall i$:
\begin{equation}
x_i \geq   k \bar{x}  + (k-1)\bar{\beta} 
\end{equation}
\underline{Example for the $\mathbf{S}-$matrix}: 
$k = 2N/(N+1)$. If $\beta_i = \bar{\beta} \forall i$, then: 
\begin{align*}
\mathbf{V}_S (\hat{\mathbf{x}}) 
= \frac{2N}{N+1} \bar{x} \mathbf{1}_N + \frac{2N}{N+1} \bar{\beta} \mathbf{1}_N 
\approx (2 \bar{x} + 2 \bar{\beta} ) \mathbf{1}_N \hspace{0.2cm} \text{ for } N \gg 1
\end{align*}


\subsubsection*{Additional constant offset $\alpha$ on the multiplexing matrix}
\label{Matrix with cte offset}
This scenario represents the case where, instead of multiplexing by the planned matrix $\mathbf{A}$, the object is actually multiplexed by this matrix plus an offset $\alpha$ that depends on the object signal. It is the proportion of the total signal from the object that contributes to the measurement when it is not expected to be. The model is: 
\begin{equation} 
\mathbf{b} \sim Poisson((\mathbf{A} + \alpha \mathbf{J}_N ) \mathbf{x} )
= Poisson( \mathbf{C} \mathbf{x})
\label{eqn: model_offset_matrix}
\end{equation}
with $\mathbf{C}=\mathbf{A}+ \alpha \mathbf{J}_N$.
The variance thus derives from the initial model formula of equation \eqref{eqn:VarianceGeneralFormula_Real}: 
\begin{equation}
\boxed{\widetilde{\mathbf{V}} = ((\mathbf{A} + \alpha \mathbf{J}_N )^{-1})\odot (\mathbf{A} + \alpha \mathbf{J}_N )^{-1})( (\mathbf{A} + \alpha \mathbf{J_N} )) \mathbf{x}}
\end{equation}
For raster-scanning, $\mathbf{C}_{rs}=\mathbf{I}_N+ \alpha \mathbf{J}_N$ and 
$\mathbf{C}_{RS}^{-1} = \mathbf{I}_N - \frac{\alpha \mathbf{J}_N}{1+\alpha N}$ (equation \eqref{eqn:Sherman_Morrison}). 
Thus,
\begin{align*}
\widetilde{\mathbf{V}}_{RS}(\hat{\mathbf{x}})  & = (\mathbf{C}_{RS}^{-1} \odot \mathbf{C}_{RS}^{-1})(\mathbf{C}_{RS}\mathbf{x}) \\
& = (\mathbf{I}_N - \frac{\alpha \mathbf{J}_N}{1+\alpha N})^{\odot 2} (\mathbf{I}_N + \alpha \mathbf{J}_N)\mathbf{x}\\
& = (\mathbf{I}_N - \frac{2\alpha}{1+\alpha N}\mathbf{I}_N +  \frac{\alpha^2}{(1+\alpha N)^2}\mathbf{J}_N)(\mathbf{I}_N+ \alpha \mathbf{J}_N)\mathbf{x}\\
& = (1 - \frac{2\alpha}{1+\alpha N})\mathbf{x} + (\alpha - \frac{2\alpha^2}{1+\alpha N} + \frac{\alpha^2+ \alpha^3 N}{(1+\alpha N)^2})N\bar{x}\mathbf{1}_N\\
& = (1 - \frac{2\alpha}{1+\alpha N})\mathbf{x} + \frac{\alpha N ((1+\alpha N)^2 -\alpha(1+\alpha N))    }{(1+\alpha N)^2}\bar{x} \\
& = (1 - \frac{2\alpha}{1+\alpha N})\mathbf{x} + \frac{\alpha N (1+\alpha N -\alpha)}{1+\alpha N}\bar{x} 
\end{align*}
and, if $\alpha \ll 1$ or $N \gg 1$: 
\begin{equation}
\boxed{\widetilde{\mathbf{V}}_{RS} (\hat{\mathbf{x}}) \approx  \mathbf{x} + \alpha N \bar{x} \mathbf{1}_N}
\end{equation}

\noindent For multiplexing with $\mathbf{A_c}$, $\mathbf{C}_{Ac}=\mathbf{A_c}+ \alpha \mathbf{J}_N$. We use the Sherman-Morrinson formula  (equation \eqref{eqn:Sherman_Morrison}) to calculate the inverse of $\mathbf{C}_A$:
\begin{align*}
\mathbf{C}_{Ac}^{-1} & = \mathbf{A_c}^{-1} - \frac{\alpha \mathbf{A_c}^{-1}\mathbf{J}_N \mathbf{A_c}^{-1}}{1+\alpha \mathbf{1}^T \mathbf{A}^{-1} \mathbf{1}} \\ 
& = \mathbf{A_c}^{-1} - \frac{\alpha k (N-2p)^2}{N(1+\alpha k (N-2p)^2)} \mathbf{A_c}^{-1}\mathbf{J}_N \\ 
& = \mathbf{A_c}^{-1} \left( \mathbf{I}_N - \frac{\alpha k q^2}{N(1+\alpha k q^2)} \mathbf{J}_N \right)
\end{align*}
with $q = (N-2p)$. We would need a further condition on the sum of the rows of $\mathbf{A}$ to be able to carry out the derivation. 
If $N \gg 1$ and $q^2$ is independent of $N$ (true for matrices $\mathbf{S}$, $\mathbf{H_1}$ and $\mathbf{C_1}$), then $\mathbf{C}_{Ac}^{-1} \approx \mathbf{A_c}^{-1}$ and: 
\begin{align*}
\widetilde{\mathbf{V}}_{Ac} (\hat{\mathbf{x}}) & \approx (\mathbf{A_c}^{-1} \odot \mathbf{A_c}^{-1}) (\mathbf{A_c} + \alpha \mathbf{J}_N) \mathbf{x} \\
& = k\bar{x}\mathbf{1}_N + \alpha \frac{k}{N} \mathbf{J}_N \mathbf{A_c}^{-1} \mathbf{J}_N  \mathbf{x} \\
& = k\bar{x}\mathbf{1}_N + \alpha \frac{k^2 q^2}{N^2} \mathbf{J}_N \mathbf{J}_N  \mathbf{x}
\end{align*}
i.e. 
\begin{equation}
\boxed{\widetilde{\mathbf{V}}_{Ac}  (\hat{\mathbf{x}}) \approx (k  + \alpha k^2 q^2) \bar{x}\mathbf{1}_N }
\end{equation}
if in addition, $\alpha \ll 1$, multiplexing with $\mathbf{A_c}$ is advantageous over raster-scanning for pixels $i$ for which: 
\begin{equation}
\boxed{x_i \geq  k \bar{x}  - N \alpha \bar{x}  }
\end{equation}

\noindent \underline{Example of the $\mathbf{S}-$ matrix}: $\mathbf{C}_S=\mathbf{S}+ \alpha \mathbf{J}_N$ and it inverse reads: 
\begin{equation}
\mathbf{C}_S^{-1} = \mathbf{S}^{-1} - \frac{\alpha \mathbf{S}^{-1}\mathbf{J}_N \mathbf{S}^{-1}}{1+\alpha \mathbf{1}^T \mathbf{S}^{-1} \mathbf{1}} = \mathbf{S}^{-1} - \frac{4 \alpha}{(N+1)(1+N+2\alpha N)} \mathbf{J}_N
\end{equation}
where we used the fact that $\mathbf{J_N}\mathbf{S}^{-1} = \mathbf{S}^{-1}\mathbf{J_N} = 2/(N+1)$ , and that the sum of all elements of $\mathbf{S}^{-1}$, i.e.
$\mathbf{1}^T \mathbf{S}^{-1} \mathbf{1}$ is $2N/(N+1) $.
Then, using equations \eqref{eqn:S.S} and \eqref{eqn:Smatrixpty} leads to:
\begin{align*}
\widetilde{\mathbf{V}}_S (\hat{\mathbf{x}}) & = (\mathbf{C}_{S}^{-1} \odot  \mathbf{C}_{S}^{-1})(\mathbf{C}_{S}\mathbf{x}) \\
& = (\mathbf{S}^{-1} - \frac{4 \alpha}{(N+1)(1+N+2\alpha N)} \mathbf{J}_N)^{\odot 2} (\mathbf{S}^{-1}+ \alpha \mathbf{J}_N)\mathbf{x}\\
& = ( \frac{4}{(N+1)^2}(1 + \frac{4 \alpha^2}{(1+N+2\alpha N)^2})\mathbf{J}_N - \frac{8 \alpha}{(N+1)(1+N+2\alpha N)})\mathbf{S}^{-1} ) (\mathbf{S} + \alpha \mathbf{J}_N) \mathbf{x} \\
& = ( ( \frac{2}{N+1} + \frac{4 \alpha N}{(N+1)^2} )(1 + \frac{4 \alpha^2}{(1+N+2\alpha N)^2}) - \frac{16 \alpha^2}{(N+1)^2 (1+N+2\alpha N)})\mathbf{J_N} \mathbf{x} \\
& - \frac{8 \alpha}{(N+1)(1+N+2\alpha N)}\mathbf{I}_N\mathbf{x} 
\end{align*}
Then, for $N\gg 1$:
\begin{align*}
\widetilde{\mathbf{V}}_S (\hat{\mathbf{x}}) & \approx ( ( \frac{2}{N} + \frac{4 \alpha N}{N^2} )(1 + \frac{4 \alpha^2}{((N+2\alpha N)^2}) - \frac{16 \alpha^2}{N^2 (N+2\alpha N)})N\bar{x}\mathbf{1}_N - \frac{8 \alpha}{N(N+2\alpha N)}\mathbf{x} \\ 
& \approx ( 2(1+2\alpha)(1 + \frac{4 \alpha^2}{N^2(1+2\alpha)^2}) - \frac{16 \alpha^2}{N^2 (1+2\alpha)})\bar{x}\mathbf{1}_N - \frac{8 \alpha}{N^2(1+2\alpha)}\mathbf{x} \\
& \approx ( 2 + 4\alpha)\bar{x}\mathbf{1}_N - \frac{8 \alpha}{N^2(1+2\alpha)}(\mathbf{x} + \bar{x}\mathbf{1}_N ) 
\end{align*}
Therefore, if  $N \gg 1$: 
\begin{equation}
\widetilde{\mathbf{V}}_S (\hat{\mathbf{x}})  \approx ( 2 + 4\alpha)\bar{x}\mathbf{1}_N  
\end{equation}
In this case, S-multiplexing is therefore advantageous for pixels $i$ for which: 
\begin{align*}
& x_i + \alpha N \bar{x} \geq  2 \bar{x}  +  4\alpha \bar{x} \hspace{0.5cm} \text{i.e.} \\
& x_i \geq  2 \bar{x}  - N  \alpha \bar{x} \\
\end{align*}

\chapter{Some facts on matrices leading to a constant MSE}
\label{sec: SI_Generalisation}

In chapter \ref{sec: SI_CasesHadaFourier}, we found that for three common multiplexing matrices, namely the $\mathbf{S}$-matrix, the $\mathbf{H1}$-matrix and the $\mathbf{W1}$- matrix, when $N \gg 1$, the estimation variance reads: $V (\hat{x}_i) \approx k \bar{x}$ on most object pixels $i$.
The $\mathbf{S}$-matrix leads to a constant variance on strictly all object pixels; while the $\mathbf{H1}$ and $\mathbf{W1}$-matrices lead to a constant variance on most object pixels, but with differences on few pixels. \\

\noindent To conclude this work, we elaborate on some common characteristics of matrices leading to a strictly constant variance ($V (\hat{x}_i) = k \bar{x}$ $\forall i$). In other words, we seek conditions on the matrix $\mathbf{A} \in \mathbb{R}_{+}^{N \times N}$ so that the estimation variance is equal to a constant times the object average $\bar{x}$, on every pixel $i$
\begin{equation}
\forall \mathbf{x} \in \mathbb{R}^{+}, \hspace{0.3cm} \mathbf{V} (\hat{\mathbf{x}})  = k \bar{x} \mathbf{1}_N 
\label{eqn:VarianceDefinepb_NecSuffConditions}
\end{equation}
where $k \in \mathbb{R}_{+}^{*}$ is some constant. \\

\noindent We remind that we consider shot-noise limited measurements $\mathbf{b} \sim Poisson(\mathbf{Ax})$, where $\mathbf{A} \in \mathbb{R}^{N \times N}$ is an invertible multiplexing matrix with coefficients $a_{ij} \in \mathbb{R}^{+}$. $\mathbf{x} \in \mathbb{R}^{+} $ is the intensity object. If the estimate is obtained via least-square estimation ($\hat{\mathbf{x}} = \mathbf{A}^{-1} \mathbf{b}$), the estimation variance reads $\mathbf{V} (\hat{\mathbf{x}}) = ( \mathbf{A}^{-1} \odot \mathbf{A}^{-1}) \mathbf{A} \mathbf{x}$.

\section{A necessary condition}

Here, $\forall \mathbf{x} \in \mathbb{R}^{+}$, we seek invertible matrices $\mathbf{A} \in \mathbb{R}_{+}^{N^2}$ such that:
\begin{align*}
\mathbf{V} (\hat{\mathbf{x}}) & = k \bar{x} \mathbf{1}_N \\
\Leftrightarrow ( \mathbf{A}^{-1} \odot \mathbf{A}^{-1}) \mathbf{A} \mathbf{x} & = \frac{k}{N} \mathbf{J}_N \mathbf{x} \\
\Leftrightarrow (\mathbf{A}^{-1})^{\odot 2} & = \frac{k}{N} \mathbf{J}_N  \mathbf{A}^{-1} \\
\Leftrightarrow   t_{ij}^2  & = \frac{k}{N} \sum_{q} t_{qj} \hspace{0.1cm} \forall q,j
\end{align*}
where $t_{ij} = (\mathbf{A}^{-1})_{ij}$ denotes the elements of $\mathbf{A}^{-1}$ and $\sum_{q} t_{qj}$ is the sum of the elements of the column $j$ of $\mathbf{A}^{-1}$. This means that all the elements of a given column $j$ of $(\mathbf{A}^{-1})^{\odot 2}$ are equal, and proportional to the sum of the elements of the column $j$ of $\mathbf{A}^{-1}$. This also means that the absolute value of all the elements of a given column $j$ of $\mathbf{A}^{-1}$ are equal. 
Defining $l_j = \sqrt{\frac{k}{N} \sum_{q} t_{qj}} > 0$ ($l_j \neq 0 $ since $\mathbf{A}$ is invertible), we have $t_{ij}^2 =  l_j^2$ and  $t_{ij} = \pm l_j= \delta_{ij} l_j$, with $\delta_{ij} = -1$ or $+1$. \\
In the matrix form, this reads: 
\begin{equation}
(\mathbf{A}^{-1})^{\odot 2}
= 
\begin{pmatrix}
l_1^2 & \cdots &l_N^2 \\
\vdots &  \vdots & \vdots \\
l_1^2  & \cdots &l_N^2
\end{pmatrix}
=
\begin{pmatrix}
t_{11}^2 & \cdots &t_{1N}^2 \\
\vdots &  \ddots & \vdots \\
t_{N1}^2  & \cdots &t_{NN}^2
\end{pmatrix}
= 
\frac{k}{N}
 \begin{pmatrix}
\sum_{i} t_{i1} & \cdots & \sum_{i} t_{iN} \\
\vdots &   & \vdots \\
\sum_{i} t_{i1}  & \cdots & \sum_{i} t_{iN}
\end{pmatrix} \\
\end{equation}
and therefore: 
\begin{equation}
\mathbf{A}^{-1} 
= 
\begin{pmatrix}
\delta_{11} l_1 & \cdots & \delta_{1N}  l_N \\
\vdots &  \ddots & \vdots \\
\delta_{1N} l_1  & \cdots & \delta_{NN} l_N
\end{pmatrix}
\label{eqn:GeneralForminvA}
\end{equation}
Again, we know that all the elements of a given column $j$ of $(\mathbf{A}^{-1})^{\odot 2}$ are equal to $k/N$ times the sum of the elements of the column $j$ of $\mathbf{A}^{-1}$. Denoting $p_j$ the number of negative elements (i.e. with $\delta_{ij} = -1$) in column $j$, and  $(N-p_j)$ the number of  positive elements (i.e. with $\delta_{ij} = +1$) in column $j$, we have, $\forall$ column $j$: 
\begin{align*}
l_j^2 & = \frac{k}{N} \left( (N-p_j)l_j - p_j l_j \right) \\
\Leftrightarrow  l_j^2 & = \frac{k}{N} ( N - 2p_j )l_j \\
\Leftrightarrow  l_j & = \frac{k}{N} ( N - 2p_j ) \text{ ( } l_j \neq 0 \text{ ) }
\end{align*}
Note that since $l_j > 0$, there must be strictly more positive than negative elements in each column of $\mathbf{A}^{-1}$:
if $N$ is odd, the maximum number of negative elements in column $j$ is $\frac{N-1}{2}$; if $N$ is even, the maximum number of negative elements in column $j$ is $\frac{N}{2} - 1$. Last, since $\mathbf{A}$ is invertible, $p_j = 0$ in one column at most, and therefore $N > 2$. \\

\noindent Overall, we showed that, $\forall \mathbf{x} \in \mathbb{R}^{+}$,  $\forall N > 2$, and for $\mathbf{A} \in \mathbb{R}_{+}^{N^2}$ invertible: 
\begin{equation}
\boxed{
\mathbf{V} (\hat{\mathbf{x}})  = k \bar{x} \mathbf{1}_N 
\Longrightarrow 
\mathbf{A}^{-1} 
= \frac{k}{N}
\begin{pmatrix}
\delta_{11} (N-2p_1) & \cdots & \delta_{1N}  (N-2p_N) \\
\vdots &  \vdots & \vdots \\
\delta_{1N} (N-2p_1)  & \cdots & \delta_{NN} (N-2p_N)
\end{pmatrix}}
\label{eqn:NecessaryConditionAinvfinal}
\end{equation} 
where $k \in \mathbb{R}^{+*}$ is some constant, $\delta_{ij} = -1$ or $+1$, and $p_j$ is the number of negative elements $\delta_{ij} = -1$ in column $j$ (with $p_j \leq \frac{N-1}{2}$ if $N$ is odd, $p_j \leq \frac{N}{2} - 1$ if $N$ even, and $p_j = 0$ in one column at most). 
The sum of the column $j$ of $\mathbf{A}^{-1}$ is $(k/N) (N - 2 p_j)^2$. In addition, no column can be a linear combinaison of the other, which limits the number of possibilities. \\

\noindent In conclusion, $\mathbf{A}^{-1}$ is therefore a matrix in which the absolute value of all elements of a given column $j$ are equal: it is equal to a number that only depends on the number of negative element of the same column. 
Note that this condition is only necessary and not sufficient. \\

\noindent \underline{Note on special matrices:}\\
The $\mathbf{S}-$matrix used throughout this study verifies \eqref{eqn:NecessaryConditionAinvfinal} with, $\forall j$, $p_j = p = \frac{N-1}{2}$ , $l_j = l = \frac{k}{N}$ and $k = \frac{2N}{N+1} \approx 2 $. \\
Note that $\mathbf{H1}$ does not does not verify \eqref{eqn:NecessaryConditionAinvfinal} and does not exactly lead to $\mathbf{V} (\hat{\mathbf{x}}) = k \bar{x} \mathbf{1}_N$, in particular due to its first row and column.

\section{Sufficient condition 1}
Here, we consider $\mathbf{A}^{-1}$ of the form of \eqref{eqn:NecessaryConditionAinvfinal}, and add the hypothesis that all $p_j$ are equal, i.e. that, $\forall j$: 
\begin{equation}
p_j  = p \text{, i.e. } l_j = \frac{k}{N} (N - 2 p)
\end{equation}
Then, we have:
\begin{equation}
\mathbf{A}^{-1} 
= \frac{k}{N} (N-2p)
\begin{pmatrix}
\delta_{11}  & \cdots & \delta_{1N}  \\
\vdots &  \ddots & \vdots \\
\delta_{1N}  & \cdots & \delta_{NN} 
\end{pmatrix} =
\frac{k}{N} (N-2p) \mathbf{\Lambda} 
\label{eqn:SufficientConditionAinv}
\end{equation}
Therefore, $\mathbf{A}^{-1}$ is proportional to a matrix $\mathbf{\Lambda}$ made of $+1$ and $-1$ signs ($|\delta_{ij}| = 1$), where there are exactly $p$ negative elements in each column. Note that many combinations of signs, in particular permutations, may be found. Note also that the sum of all columns of $\mathbf{A}^{-1}$ is a constant: 
\begin{equation}
\mathbf{J}_N \mathbf{A}^{-1} = \frac{k}{N} (N - 2p)^2  \mathbf{J}_N
\end{equation}
and therefore the sum of all columns of $\mathbf{A}$ is also a constant:
\begin{equation}
\mathbf{J}_N \mathbf{A} = \frac{N}{k(N - 2p)^2} \mathbf{J}_N
\label{eqn:sumAconstant}
\end{equation}
Then, it is easy to verify that: 
\begin{align*}
\mathbf{V} (\hat{\mathbf{x}}) = (\mathbf{A}^{-1})^{\odot 2}  \mathbf{Ax} = \frac{k^2}{N^2}(N - 2p)^2 \mathbf{J}_N \mathbf{Ax} = \frac{k^2}{N^2}(N - 2p)^2  \frac{N}{k(N - 2p)^2} \mathbf{J}_N \mathbf{x}  = \frac{k}{N} \mathbf{J}_N \mathbf{x} = k \bar{x} \mathbf{1}_N
\end{align*}
Therefore, we can write, $\forall \mathbf{x} \in \mathbb{R}^{+}$,  $\forall N > 2$, and for $\mathbf{A} \in \mathbb{R}_{+}^{N^2}$ invertible: 
\begin{equation}
\boxed{\mathbf{A}^{-1} = \frac{k}{N} (N-2p) \mathbf{\Lambda}
\Longrightarrow 
\mathbf{V}(\hat{\mathbf{x}}) = k \bar{x} \mathbf{1}_N}
\label{eqn:SufficientConditionAinvfinal}
\end{equation}
where $k \in \mathbb{R}_{+}^{*}$ is some constant and $\mathbf{\Lambda}$ is a matrix with elements $\delta_{ij} = \pm 1$, in which each column contains $p$ negative elements. \\

\noindent \underline{Note on special matrices:}\\
The $\mathbf{S}-$matrix used throughout this study verifies \eqref{eqn:SufficientConditionAinvfinal} with $ p = \frac{N-1}{2}$ , $l = \frac{k}{N}$ and $k = \frac{2N}{N+1} \approx 2 $.  \\
Note that $\mathbf{H1}$ does not does not verify \eqref{eqn:SufficientConditionAinvfinal} in particular due to its first row and column.

\section{Sufficient condition 2}

Independently from the above results, another sufficient condition is straightforward. Since $\mathbf{V}(\hat{\mathbf{x}}) = (\mathbf{A}^{-1} \odot \mathbf{A}^{-1}) \mathbf{A} \mathbf{x}$, an invertible matrix $\mathbf{A} \in \mathbb{R}_{+}^{N^2}$ in which the sum of its columns is constant and the Hadamard product of its inverse is proportional to $\mathbf{J}_N$ leads to a variance equal to a constant times the object average $\bar{x}$: 
\begin{equation}
\left\{
    \begin{array}{ll}
        \mathbf{A}^{-1} \odot \mathbf{A}^{-1} = \frac{\theta_1}{N} \mathbf{J}_N \\
       \mathbf{J}_N  \mathbf{A} = \theta_2 \mathbf{J}_N \\
    \end{array}
\right.
\Longrightarrow 
\mathbf{V}(\hat{\mathbf{x}}) = \theta_1 \theta_2 \bar{x} \mathbf{1}_N = k \bar{x} \mathbf{1}_N
\end{equation}
where $\theta_1 \in \mathbb{R}^{+*}$ and $\theta_2 \in \mathbb{R}^{+}$ are some constants. \\ 

\section*{Conclusion}
\label{sec: SI_Conclusion}

\vspace{1cm}

In this study, we provided a detailed methodology and derivations to analyse the SNR of single-pixel detection multiplexing under photon noise. To assess the theoretical SNR performances, we derived the SNR associated with some types of single-pixel detection multiplexing, for three measurements schemes (One-step, Two-step, Dual-detection), both in the general case and for Hadamard-based and Cosine-based positive-multiplexing. We showed that in the particular case Hadamard-based and Cosine-based positive-multiplexing, the MSE is approximately constant on most object pixels. This implies that, as compared to raster scanning, such types of single-pixel detection multiplexing do not systematically improve the SNR, but only improve it on object pixels at least $k$ times brighter than the object mean signal $\bar{x}$. Since other multiplexing matrices with the same property may exist, we derived some conditions on such matrices. We also provided a detailed robustness study that showed that in most of the studied cases, positive-multiplexing is more robust that raster-scanning to additional perturbations, unless the perturbation itself experiences multiplexing.
The practical implications of the present theoretical results - which are crucial when choosing an optical system or an acquisition strategy - will be studied in a forthcoming publication. \\ \\ \\ \\ \\ \\ \\

\noindent \textbf{Funding}\\
C. S. has received funding from the H2020 Marie Skłodowska-Curie Actions (713750). This research has received funding from EU ICT-36-2020RIA CRIMSON, Agence Nationale de la Recherche  (ANR-21-ESRS-0002 IDEC), Centre National de la Recherche Scientifique, Aix-Marseille University. \\

\noindent \textbf{Acknowledgements}\\
The authors thank Simon Labouesse, Siddharth Sivankutty, Philippe Réfrégier, Laurent Jacques, Randy A. Bartels, Marc Allain, Anne Sentenac, Sandro Heuke and Luis Arturo Aleman Castaneda for fruitful scientific discussions. \\ 

\noindent \textbf{Authors Contributions}\\
C.S. performed the calculations and simulations, and wrote the paper. 
All authors contributed to the scientific discussion and revision of the paper.\\

\noindent \textbf{Competing interests}\\
The authors declare no conflict of interest.\\







\cleardoublepage


\bigskip
\bibliography{CitationsMultiplexingpaper}
\bibliographystyle{unsrt}

\end{document}